\documentclass[journal]{IEEEtran}
\makeatletter
\let\MYcaption\@makecaption
\makeatother

\usepackage{cite}
\usepackage{tikz}
\usepackage[percent]{overpic}
\usepackage{amsmath,amsfonts,amssymb}
\usepackage{graphicx}
\usepackage{overpic}
\usepackage{tikz}
\usepackage{xcolor}
\usepackage{mathdots}
\usepackage{pict2e}
\usepackage{bbm}
\usepackage{mathtools, cuted}
\usepackage{amssymb,bm}
\usepackage{ragged2e}
\usepackage{overpic}
\usepackage{float}
\usepackage{color}
\usepackage[super]{nth}
\usepackage{graphicx}
\usepackage{textcomp}
\usepackage{tikz}
\usetikzlibrary{shapes.geometric, arrows, automata, fit, positioning, quotes, shapes.multipart}
\usepackage{array}
\usepackage[crop=pdfcrop]{pstool}
\usepackage{cancel}
\usepackage{dblfloatfix} 
\usepackage{tabularx}
\usepackage[usestackEOL]{stackengine}
\usepackage{blindtext}
\usepackage{float}
\usepackage{siunitx}
\usepackage{setspace}
\usepackage{algorithmic}
\usepackage{multirow}
\usepackage{psfrag}
\usepackage{verbatim}
\usepackage{url}\usepackage{gensymb}
\usepackage{balance}
\usepackage{subcaption}

\newcommand{\bbmatrix}{\begin{bmatrix}}
	\newcommand{\ebmatrix}{\end{bmatrix}}

\definecolor{burntorange}{rgb}{0.8, 0.28, 0.0}
\definecolor{myGreen}{rgb}{0.0, 0.5, 0.0}
\definecolor{amber}{rgb}{0.8, 0.28, 0.0}
\definecolor{ceruleanblue}{rgb}{0.16, 0.28, 0.75}

\usepackage[normalem]{ulem}

\definecolor{ao}{rgb}{0.0, 0.5, 0.0}
\definecolor{cobalt}{rgb}{0.0, 0.28, 0.67}
\definecolor{amber}{rgb}{0.8, 0.36, 0.27}

\begin{document}

\title{A Miniaturized Broadband 1-Bit Coding Reconfigurable Intelligent Surface for NLOS UE Localization and Uplink Communication}

\author{Khagendra~Joshi,~\IEEEmembership{Graduate~Student~Member,~IEEE,}
Deepak~Kumar~Sahoo,~\IEEEmembership{Graduate~Student~Member,~IEEE,}
Kamalesh~Kumar~K,
Debidas~Kundu,~\IEEEmembership{Senior Member,~IEEE,}
Vivek~A.~Bohara,~\IEEEmembership{Senior Member,~IEEE,}
and~Amalendu~Patnaik,~\IEEEmembership{Senior~Member,~IEEE}

\thanks{K. Joshi and V. A. Bohara are with the Department of Electronics and Communication Engineering, Indraprastha Institute of Information Technology Delhi, Delhi, India. Email: khagendra@iiitd.ac.in
}
\thanks{D. K. Sahoo and A. Patnaik are with the Department of Electronics and Communication Engineering, Indian Institute of Technology Roorkee, Uttarakhand, India. E-mail: deepak.ece@sric.iitr.ac.in
}      
\thanks{Kamalesh. K. K. is with the Department of Electronics, at the Vellore Institute of Technology, Andhra Pradesh, India. 
}
\thanks{D. Kundu is with the Department of Electrical Engineering, Indian Institute of Technology Delhi, Delhi, India. Email: debidask@ee.iitd.ac.in}
}

\markboth{Manuscript Draft}%
{Shell \MakeLowercase{\textit{et al.}}: Bare Demo of IEEEtran.cls for IEEE Journals}

\maketitle

\begin{abstract}
In this paper, a broadband 1-bit coding metasurface-based reconfigurable intelligent surface (RIS) is presented. The unit cell of the metasurface consists of a wide dipole modified with interdigital capacitors and loaded with an SMP 1340-040LF PIN diode. The proposed element offers cell miniaturization and a stable angular response. A phase difference of 180$\degree \pm$ 30$\degree$ is achieved for a frequency range of 4.85-6.05 GHz between the ON and OFF states for the normal incidence of the TE polarized wave, whereas it provides a fairly stable response with reflection loss of less than 3 dB and phase difference of 180$\degree$ $\pm$ 50$\degree$ for oblique incidence up to 45$\degree$. The RF is isolated from the DC on the bias lines using properly designed butterfly-shaped radial stubs. Using this unit cell, a prototype with an array of 16 $\times$ 10 elements is constructed. A low-cost microcontroller-based control circuit is designed, which can be plugged-in for biasing the PIN diodes of such array. The theoretically calculated and full-wave simulated radiation patterns of the array are validated using experiments inside anechoic chamber. Furthermore, the capability of the RIS for non-line of sight (NLOS) user equipment (UE) localization and robust uplink communication is demonstrated using LTE communication framework. This shows great potential of our RIS for applications, such as in unmanned aerial vehicle (UAV) localization and its uplink communication at NLOS or extended range. 
\end{abstract}

\begin{IEEEkeywords}
Beamforming, intelligent, metasurface, reconfigurable, uplink
\end{IEEEkeywords}

\maketitle

\section{Introduction}

\IEEEPARstart{R}{econfigurable} intelligent surface (RIS) is gradually proving to be a promising technology for future wireless communication systems. The main component of RIS is the metasurface unit cell which can manipulate the magnitude and/or phase of EM waves \cite{yu2014flat}. The unit cell of the metasurface can be loaded with tunable elements, such as PIN diode, varactor diode, and liquid crystals, to flexibly vary the magnitude and/ or phase with bias control to achieve dynamic beamforming useful for wireless communications \cite{feng2023reconfigurable}. Many key performance indices (KPIs) are observed to be improved in a RIS-aided wireless channel \cite{9140329, dai2021wireless}. In the literature, most of the RIS are designed using metasurfaces with quantized phase states, popularly known as coding metasurfaces \cite{cui2014coding}, to reduce hardware cost and complexity and improve energy efficiency. 

Numerous studies have focused on PIN diode-loaded 1-bit coding metasurface-based RIS to explore its wide range of useful applications \cite{zhang2022dual, zhang2018space, yang2016programmable, xu2024low, trichopoulos2022design}. While 1-bit designs inherently produce symmetric grating lobes under plane wave incidence \cite{narayanan2024optimum}, their cost effectiveness and simplicity lower reflection loss and wider bandwidth, while being effective in many practical scenarios, make them an attractive choice. As a result, research in this area is still dominated by 1-bit configurations compared to multi-bit counterparts despite the superior beamforming performance in terms of reduced side-lobe levels, lower beam-squint error offered by the latter.

As the demand for more reliable connections and higher data rates have increased, flexible beamforming using broadband low-cost RIS becomes crucial. The most common techniques to increase the bandwidth of 1-bit coding metasurface are to use a thick and low-permittivity substrate to reduce the Q-factor of the resonance \cite{gauthier1997microstrip, katehi1983effect}. As a result, low-cost, easily available FR4 substrate ($\epsilon_{r} = 4.3$) is often ignored when designing broadband 1-bit coding RIS. Another drawback of the FR4 material is its high dielectric loss ($\tan \delta = 0.025$) which leads to a significant reflection loss from RIS when not carefully designed. On the contrary, a high permittivity of the substrate shifts the resonant frequency towards lower frequencies, thereby providing unit cell miniaturization. Generally, the unit cell periodicity of $\lambda/2$ makes the array large and space-consuming for a larger number of elements. Although 1-bit broadband RISs have been presented in the literature recently \cite{han2019wideband, shi2024ultra, li2024ultrawideband, wang2024wideband}, miniaturized broadband unit cell is rarely reported. 

Recently, RIS-based user equipment (UE) localization \cite{yin2024uecontrolled} has shown promise for various use cases, such as the industrial internet of things (IIoT), unmanned aerial vehicles (UAVs), and robotic arms. Moreover, they can be used for energy-efficient uplink communications required for the above applications \cite{ liu2021compact, william2025energy}. However, most reported indoor and outdoor demonstrations of RIS-based communications focus on downlink, either using BS-controlled RIS \cite{tang2020wireless, wang2022reconfigurable} or self-adaptive RIS \cite{albanese2022marisa, yang2025adaptively}.

\begin{table*}
\renewcommand{\arraystretch}{1.5}
\centering
\caption{Summary of the Related Works on Wideband 1-bit Coding RIS}
\resizebox{0.95\textwidth}{!}{%
\begin{tabular}{|c|c|c|c|c|c|c|c|} 
\hline
Reference & Operating Band (FBW \%) & Miniaturized Unit Cell & Periodicity ($\lambda_{0}$) & Metasurface Substrate & Array Size (Cells) & Steerable Range & UE Localization + Uplink Comm. \\ [1ex] \hline\hline

\cite{trichopoulos2022design} & 5.65--5.95 GHz (5.17\%) & No & 0.5$\lambda_0$ &  Rogers RT 6002 & 16$\times$10 & $\pm60^\circ$ & No \\ \hline

\cite{han2019wideband} & 4.7--5.3 GHz (12\%) & No & 0.5$\lambda_0$ & F4B & 12$\times$12 & $\pm50^\circ$ & No \\ \hline

\cite{shi2024ultra} & 2.75--6 GHz (74.3\%) & No & 0.26$\lambda_0$ & F4B & 20$\times$20& $\pm60^\circ$ & No \\ \hline

\cite{li2024ultrawideband} & 6.75--11.25 GHz (50\%) & No & 0.3$\lambda_0$ & F4B & 16$\times$16 & $\pm60^\circ$ & No \\ \hline

\cite{wang2024wideband} & 22.7--30.5 GHz (29.3\%) & No & 0.34$\lambda_0$ & Rogers RT 5880 & 20$\times$20 & $\pm50^\circ$ & No \\ \hline

\textbf{This work} & \textbf{4.85--6.05 GHz (22\%)} & \textbf{Yes} & \textbf{0.29$\lambda_0$} & \textbf{FR4}  & \textbf{16$\times$10} & \textbf{$\pm\text{45}^\circ$} & \textbf{Yes} \\ [0.5ex] \hline
\multicolumn{8}{l}{FBW: Fractional bandwidth} \\
\end{tabular}%
}
\end{table*}

In this paper, a broadband RIS with a miniaturized unit cell is presented. It provides $22 \%$ fractional bandwidth, whereas the periodicity of the unit cell is 0.29$\lambda$ (16 mm) at the center frequency 5.45 GHz. The low-cost FR4 substrate is used for this design, and special attention is paid to keep the reflection loss within the comparable range obtained by the other reported works. The proposed unit cell is used to construct an RIS prototype of 16 $\times$ 10 array. The radiation patterns of the RIS are tested inside the anechoic chamber. A summary of the state-of-the-art broadband 1-bit RIS and the proposed RIS, covering parameters such as unit cell size, substrate type, operating band, and steerable range, is provided in Table I. It can be observed that the unit cell is comparatively compact and low-cost. The unit cell provides quite stable magnitude ($|\Gamma|$ within -3 dB) and phase (difference within 180$\degree \pm$ 50$\degree$) response for oblique incidence up to 45$\degree$. It also offers a satisfactory beam steering range of $\pm$ 45$\degree$. Furthermore, the RIS is used to experimentally demonstrate the successful detection of moving UE and its uplink communications \cite{hao2024uplink, park2022reconfigurable} with dynamic beamforming in real time. To autotrack the UE position, RIS sequentially applies a set of predefined angular phase configurations, and the eNB measures the corresponding received signal power for each configuration. The phase configuration that yields the maximum received power is selected and fixed at the RIS, thereby establishing RIS-assisted non-line-of-sight (NLOS) uplink communication. The performance of RIS in improving received signal strength, achieving wideband signal-to-interference-plus-noise ratio (SINR), and maintaining stable throughput with low block error rate (BLER) is demonstrated using both theoretical and experimental results. From the results, it becomes evident that RIS helps reduce the detection complexity of  UE, and its directed beams make uplink communication better and more efficient.

\section{Unit Cell Design}

\subsection{Stack Up Details} 
\begin{figure}[!t]
	\centering
	\begin{overpic}[width=0.5\textwidth,  grid=false, tics=5, trim={0cm 0cm 0cm 0cm}, clip]{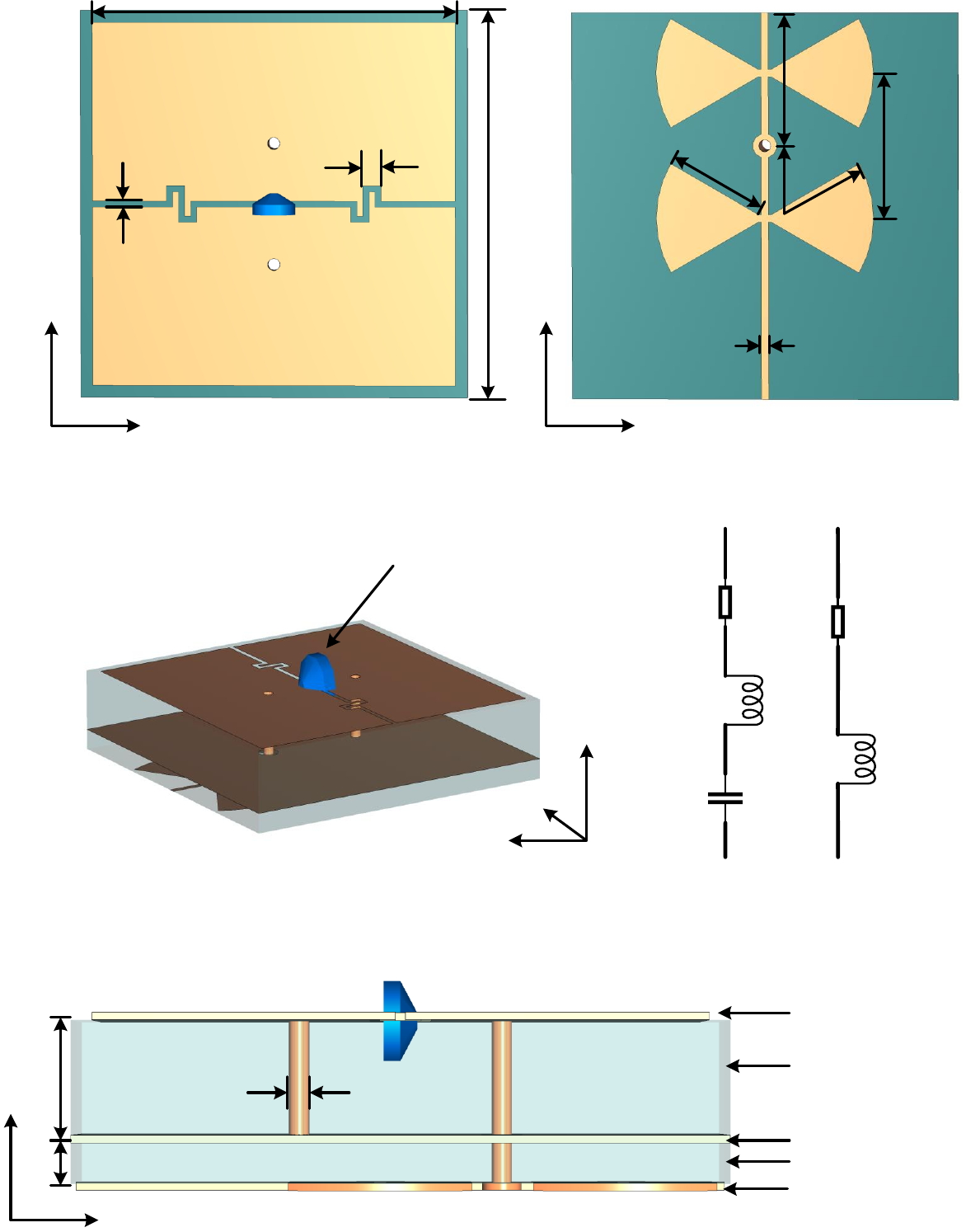}
            \put(11.5,64.75){\scriptsize X}
            \put(3.4,74.25){\scriptsize Y}
            \put(29,88.5){\scriptsize $w_{3}$}
            \put(9,87.5){\scriptsize $w_{2}$}
            \put (40,83){\scriptsize $p$}
            \put (21.0,97.65){\scriptsize $w_{1}$}
            \linethickness{0.5pt}
            \put(36,68.75){\vector(0,1){29.5}}
            \put(36,98.25){\vector(0,-1){29.5}}
            \put(34.75,68.75){\line(1,0){2.5}}
            \put(34.75,98.25){\line(1,0){2.5}}
            \put(33.5,83){\scriptsize $l_{1}$}
            \put(21.75, 62){\scriptsize (a)}
            \put(51.5,64.75){\scriptsize X}
            \put(43.55,74.25){\scriptsize Y}
            \put(72.25,87.5){\scriptsize $l_{5}$}
            \put(64,93){\scriptsize $l_{2}$}
            \put(58,86.5){\scriptsize $l_{3}$}
            \put(64,84.5){\rotatebox{30}{\scriptsize $l_{3}+l_{4}$}}
            \put(65,71.5){\scriptsize $w_{4}$}
            \put(61.75, 62){\scriptsize (b)}
            \put(42.25,34.5){\scriptsize X}
            \put(39.5,31.25){\scriptsize Y}
            \put(47,39.75){\scriptsize Z}
            \put(32.25,54){\scriptsize PIN diode}
            \put(21.75, 27){\scriptsize (c)}
            \put(59.75,50.25){\scriptsize $R_{\text{off}}$}
            \put(62.25,42.5){\scriptsize $L_{\text{off}}$}
            \put(60.75,34.5){\scriptsize $C_{\text{off}}$}
            \put(57.25,57.75){\scriptsize Off}
            \put(68.75,48.65){\scriptsize $R_{\text{on}}$}
            \put(71.5,37.75){\scriptsize $L_{\text{on}}$}
            \put(66.5,57.75){\scriptsize On}
            \put(61.75, 27){\scriptsize (d)}
            \put(64.5,17){\scriptsize Patch}
            \put(64.5,12.75){\scriptsize FR4}
            \put(64.5,6.75){\scriptsize Ground}
            \put(64.5,5){\scriptsize FR4}
            \put(64.5,2.75){\scriptsize Bias}
            \put(23,13.5){\scriptsize $d_{1}$}
            \put(5.5,5){\scriptsize $h_{2}$}
            \put(5.5,12){\scriptsize $h_{1}$}
            \put(0.5,10){\scriptsize Z}
            \put(8,0){\scriptsize Y}
            \put(41.75, -2){\scriptsize (e)}
	\end{overpic}
	\vspace{3pt}
	\caption{Unit cell geometry of the coding metasurface. (a)-(c) Top view, bottom view, and perspective view of the proposed unit cell, respectively. (d) Equivalent circuit model of the PIN diode. (e) Side view of substrate and copper layer stackup. Design parameters in mm: $p=16$, $l_{1}=w_{1}=15$, $w_{2}=0.25$, $w_{3}=0.75$, $w_{4}=0.3$, $l_{2}=5.5$, $l_{3}=4.5$, $l_{4}=3$, $l_{5}=6$, $d_{1}=0.5$, $h_{1}=2.4$, $h_{2}=0.8$.}
    \label{Fig. unit cell}
\end{figure}

The proposed unit cell is shown in Fig.~\ref{Fig. unit cell}. It consists of a wide dipole element, modified with interdigital capacitance and loaded with a PIN diode on the top layer. A Skyworks SMP1340-040LF PIN diode \cite{SMP1340}, which exhibits good performance at frequencies up to 6 GHz, is integrated into each unit cell. Furthermore, the unit cell includes a ground plane sandwiched between two FR4 substrate layers ($\epsilon_{r}=4.3$ and $\tan \delta=0.025$) with thickness $h_{1}$ and $h_{2}$. To achieve beam steering along the $\phi=0^{\circ}$ plane with incidence of a y-polarized plane wave, the PIN diode is connected along the y-direction. The cathode of the PIN diode is connected to the ground layer using a via, whereas the anode of the PIN diode is connected to the bias line at the bottom layer using another via bypassing the ground. The equivalent circuit models for the ON (state 1) and OFF (state 0) conditions of the PIN diodes are illustrated in Fig.~\ref{Fig. unit cell}(d), where $R_{\text{on}} = 1 \, \Omega$, $L_{\text{on}} = 0.45\,\text{nH}$, $R_{\text{off}} = 10 \, \Omega$, $L_{\text{off}} = 0.45\,\text{nH}$, $C_{\text{off}} = 0.16\,\text{pF}$ \cite{pin_diode_equivalent}.  

\begin{figure}[!t]
	\hspace*{0.35cm} 
	\begin{overpic}[width=0.45\textwidth,  grid=false, tics=5, trim={0cm 0cm 0cm 0cm}, clip]{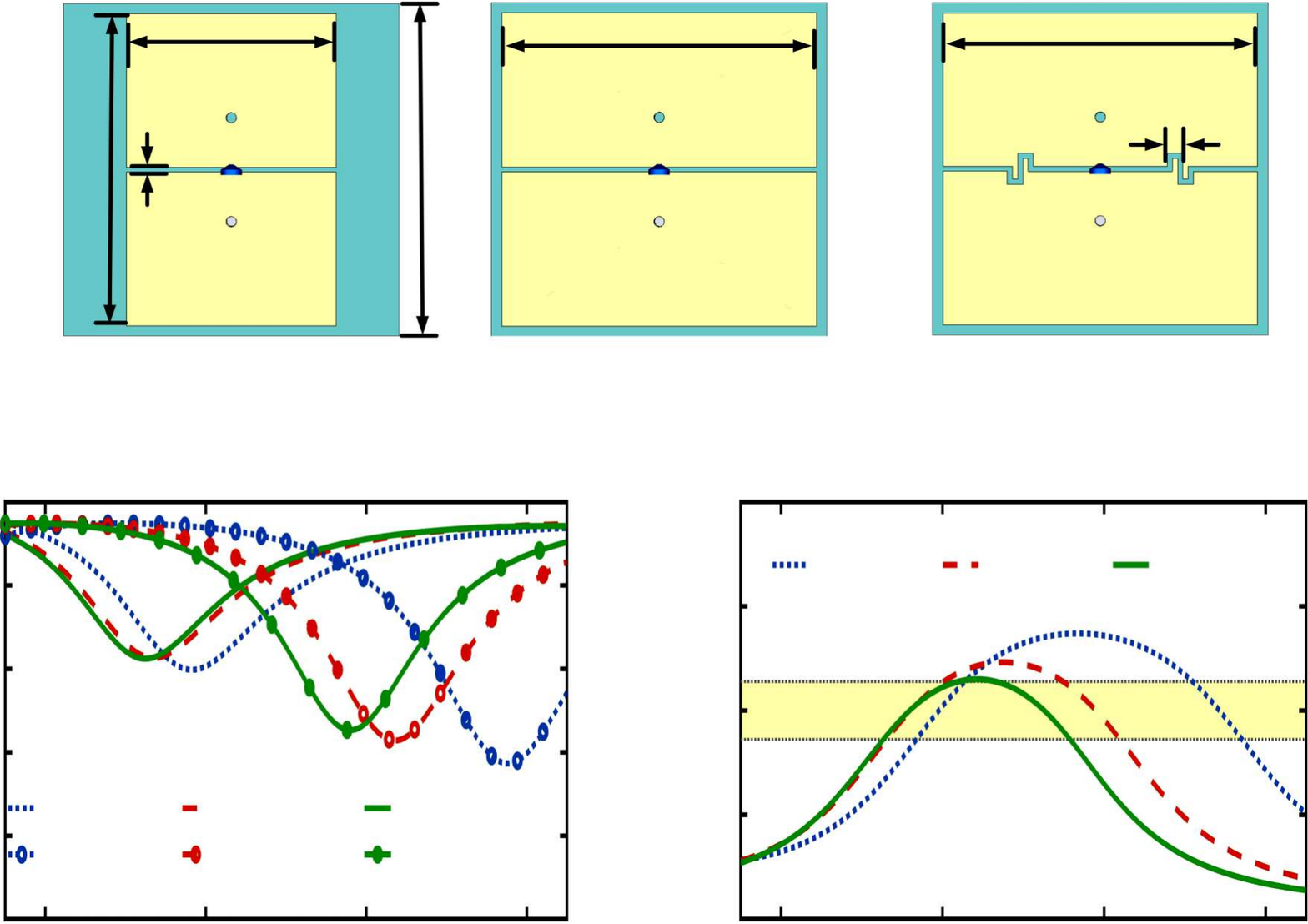}
    \put(48.5, 37.25){\scriptsize (a)}
    \put(13.5, 42.25){\scriptsize Stage I}
    \put(45.5, 42.25){\scriptsize Stage II}
    \put(80, 42.25){\scriptsize Stage III}
    \put(5.5, 57.25){\tiny $l_1$}
    \put(12, 59){\tiny $w_2$}
    \put(15.5, 68.65){\tiny $w_1$}
    \put(48.75, 68.5){\tiny $w_1$}
    \put(82, 68.5){\tiny $w_1$}
    \put(29.75, 57.25){\tiny $p$}
    \put(91.35, 60.75){\tiny $w_3$}
    \put(20, -10){\scriptsize (b)}
    \put(11.5, -6){\scriptsize Frequency (GHz)}
    \put(1.5, -2.25){\scriptsize 4.25}
    \put(13.75, -2.25){\scriptsize 5.25}
    \put(25.75, -2.25){\scriptsize 6.25}
    \put(38, -2.25){\scriptsize 7.25}
    \put(-2, 0){\scriptsize -5}
    \put(-2, 6){\scriptsize -4}
    \put(-2, 12.35){\scriptsize -3}
    \put(-2, 18.5){\scriptsize -2}
    \put(-2, 25){\scriptsize -1}
    \put(-1.25, 31){\scriptsize 0}
	\put(-6.25, 11){\rotatebox{90}{\scriptsize $|\Gamma|$ (dB)}}
    \put(3, 8.25){\tiny Stage I$^{\text{OFF}}$}
    \put(3, 4.65){\tiny Stage I$^{\text{ON}}$}
    \put(15.5, 8.25){\tiny Stage II$^{\text{OFF}}$}
    \put(15.5, 4.65){\tiny Stage II$^{\text{ON}}$}
    \put(30.25, 8.25){\tiny Stage III$^{\text{OFF}}$}
    \put(30.25, 4.65){\tiny Stage III$^{\text{ON}}$}
    \put(76.5, -10){\scriptsize (c)}
    \put(68, -6){\scriptsize Frequency (GHz)}
    \put(57, -2.25){\scriptsize 4.25}
    \put(69.25, -2.25){\scriptsize 5.25}
    \put(81.25, -2.25){\scriptsize 6.25}
    \put(93.65, -2.25){\scriptsize 7.25}
    \put(54.75, 0){\scriptsize 0}
    \put(53.2, 7.6){\scriptsize 90}
    \put(51.65, 15.35){\scriptsize 180}
    \put(51.65, 23.25){\scriptsize 270}
    \put(51.65, 31.5){\scriptsize 360}
	\put(47.85, 8.5){\rotatebox{90}{\scriptsize $\Delta \angle \Gamma$ (Deg)}}
    \put(62.25, 26.75){\tiny Stage I}
    \put(75.5, 26.75){\tiny Stage II}
    \put(88.3, 26.75){\tiny Stage III}
	\end{overpic}
	\vspace{25pt}
	\caption{(a) Evolution stages for unit cell miniaturization. Stage I: wide dipole ($w_1 = 12$, $w_2 = 0.25$), stage II: wider dipole ($w_1 = 15$, $w_2 = 0.25$), stage III: wider dipole with interdigital capacitor ($w_1 = 15$, $w_2 = 0.25$, $w_3 = 0.75$) (Unit: mm). All other parameters are kept constant during the evolution stages. (b) Reflection loss for OFF and ON states in the evolution stages. (c) Phase difference between the OFF and ON states during the three stages.}
    \label{Fig. interdigital logic}
\end{figure}

\begin{figure}[!t]
	\centering
	\begin{overpic}[width=0.45\textwidth,  grid=false, tics=5, trim={0cm 0cm 0cm 0cm}, clip]{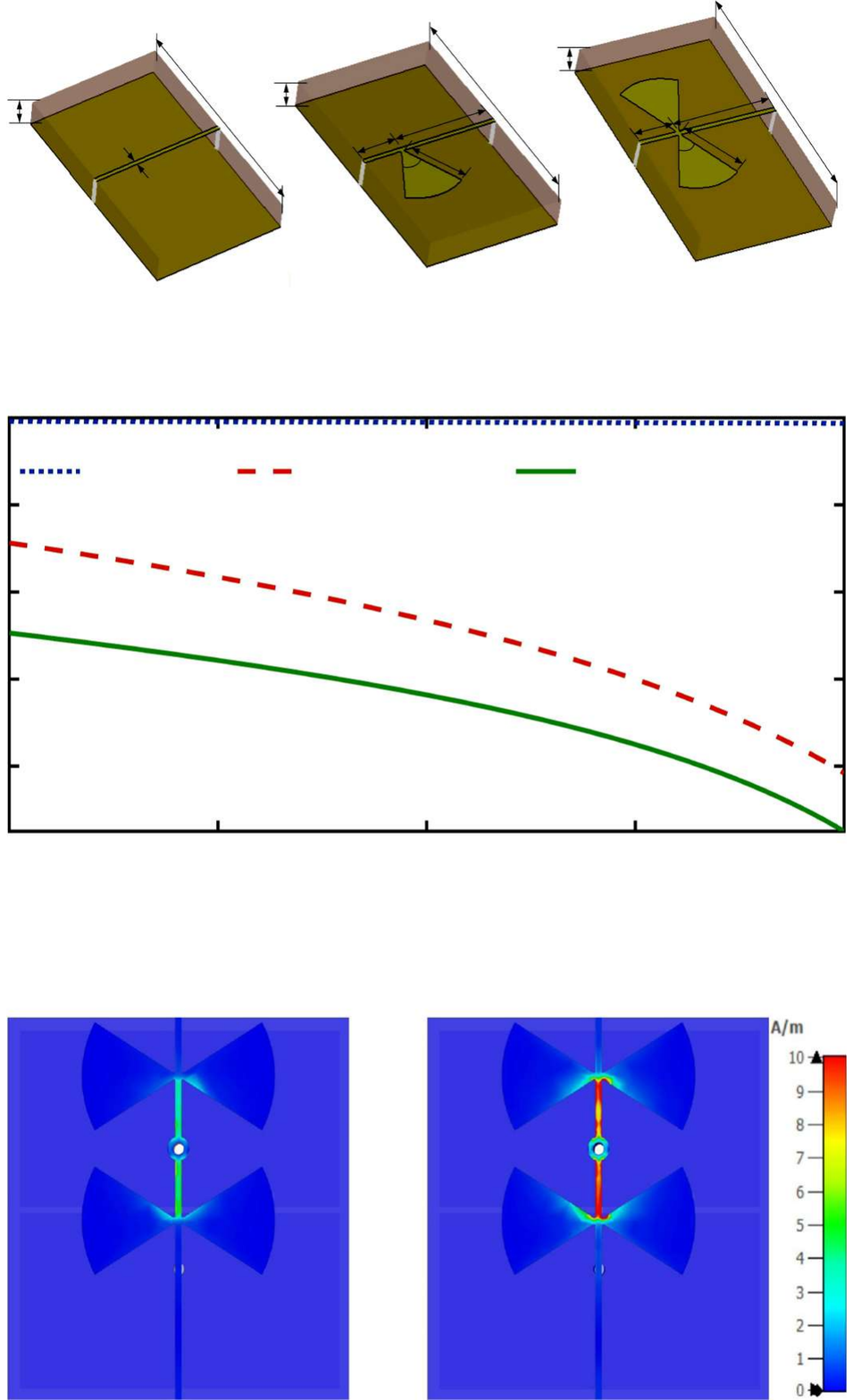}
    \put(33.75, 73.5){\scriptsize (a)}
    \put(15.75, 77){\scriptsize I}
    \put(34.25, 77){\scriptsize II}
    \put(54, 77){\scriptsize III}
    \put(3.35, 91.65){\tiny $h_2$}
    \put(22.35, 93){\tiny $h_2$}
    \put(42.65, 95.45){\tiny $h_2$}
    \put(14.25, 89.25){\tiny $w_2$}
    \put(21.0, 91.0){\tiny $l_1$}
    \put(40.5, 92){\tiny $l_1$}
    \put(60.65, 92.5){\tiny $l_1$}
    \put(8.25, 84){\tiny Port$_1$}
    \put(18.5, 88){\tiny Port$_2$}
    \put(27.25, 85.25){\tiny Port$_1$}
    \put(38.5, 88.25){\tiny Port$_2$}
    \put(46.75, 86.5){\tiny Port$_1$}
    \put(58.25, 88.9){\tiny Port$_2$}
    \put(30.15, 90.250){\tiny $l_4$}
    \put(35, 91.6){\tiny $l_6$}
    \put(35.6, 88.5){\tiny $l_3$}
    \put(50, 91.25){\tiny $l_4$}
    \put(55, 92.5){\tiny $l_6$}
    \put(55.6, 89.5){\tiny $l_3$}
    \put(34.35, 86.35){\scriptsize $\theta_{1}$}
    \put(33.75, 32){\scriptsize (b)}
    \put(28.5, 35.5){\scriptsize Frequency (GHz)}
    \put(4.5, 38.5){\scriptsize 4.5}
    \put(19.25, 38.5){\scriptsize 5.0}
    \put(34, 38.5){\scriptsize 5.5}
    \put(48.75, 38.5){\scriptsize 6.0}
    \put(63.75, 38.5){\scriptsize 6.5}
    \put(2.25, 40){\scriptsize -25}
    \put(2.25, 44.5){\scriptsize -20}
    \put(2.25, 50.75){\scriptsize -15}
    \put(2.25, 57){\scriptsize -10}
    \put(2.25, 63.25){\scriptsize -05}
    \put(2.95, 68.9){\scriptsize 00}
    \put(-1, 50){\rotatebox{90}{\scriptsize $|\text{S}_{\text{21}}|$ (dB)}}
    \put(11.5, 65.75){\scriptsize No Stub}
    \put(26.5, 65.75){\scriptsize Single Stub}
    \put(46.5, 65.75){\scriptsize Butterfly Stub}
    \put(33.75, -7){\scriptsize (c)}
    \put(15.8, -3){\scriptsize OFF}
    \put(46.1, -3){\scriptsize ON}
	\end{overpic}
	\vspace{27.5pt}
	\caption{Geometries for the study of the three biasing cases. Parameters in mm are $l_{3}=4.5$, $l_{4}=3$, $l_{6}=7.5$, $h_{2}=0.8$ and $\theta_{1}=30\degree$. (b) Simulated $|S_{21}|$ for the three biasing configurations. (c) Surface current density on the bottom layer of the proposed unit cell at 5.5 GHz for OFF and ON states of the PIN diode.}
    \label{Fig. surface current}
\end{figure}

\subsection{Miniaturization Strategy} A comprehensive parametric study of the unit cell design is carried out, with an emphasis to achieve an optimized design for IEEE 802.11a wireless local area network (WLAN) applications (5.15-5.825 GHz). The objective of the study is to ensure that the unit cell in the ON and OFF state of the PIN diode maintains a favorable reflection coefficient magnitude ($|\Gamma|$) and achieves a phase difference of $180^{\circ}\pm 30^{\circ}$ for y- (or TE-) polarized incidence over a wider frequency range of 5 to 6 GHz. In Fig.~\ref{Fig. unit cell}, the optimal design parameters of the unit cell are displayed. The logic of introducing an interdigital capacitor to the wide dipole element is explained with the help of Fig.~\ref{Fig. interdigital logic}. It can be observed that a dipole with optimum length ($l_1$) and gap ($w_2$), but increased width ($w_1$), can shift the lower cut of frequency of the phase difference as the OFF and ON state resonances are shifted due to an increase in gap-capacitance. However, the phase difference exceeds the desired range, and the reflection loss remains higher for a wider dipole. In contrast, the interdigital capacitor with the same ($w_1$) introduces a larger capacitance, which, when coupled with the wider dipole, significantly lowers the resonant Q-factor by counteracting the high dielectric loss of the FR4 substrate. As a result, keeping the periodicity still at $p =$ 16 mm, the lower cutoff frequency is reduced, thereby the electrical periodicity is reduced to $0.29\lambda_0$, and broadband phase linearity and low reflection loss are achieved. 

\subsection{RF-DC Isolation} An important aspect of the RIS unit cell design is the effective isolation of RF or alternating current (AC) from direct current (DC) on the bias lines. Otherwise, the AC behavior of the unit cell will be bias-line dependent, as well as the AC will flow to the DC supply and may damage it. In our design, we have used two butterfly-shaped radial stubs at the two sides of the non-grounded via as shown in Fig.~\ref{Fig. unit cell}(b). The distance from the end of the via to the radial stubs is $\lambda_{g}/4$ so that the radial stubs that effectively short-circuit the AC on the bias line, become open-circuited at the end of the via. An S-parameter analysis is carried out as shown in Fig.~\ref{Fig. surface current} to quantitatively demonstrate the effectiveness of the butterfly stub in RF choking. Three cases, that are bias line with no radial stub, with a single stub, and with a butterfly stub, are studied. Two discrete ports with port impedance of 141 $\Omega$ are assigned at the two ends of the bias line of width $w_{4}=0.3$ mm on $h_{2}=0.8$ mm thick grounded FR4 substrate. It can be observed that the $|S_{21}|$ at the desired frequency range is the lowest ($<-12$ dB) when the butterfly stub is used. For qualitative demonstration, the AC surface current densities on the bias line at 5.5 GHz are plotted for the ON and OFF states of the PIN diode in Fig.~\ref{Fig. surface current}(c). It can be observed that the AC is high from the via till the radial stubs, and it is significantly less after the stubs. This suggests that the RF is efficiently choked on the bias line due to the properly designed stubs. 

\subsection{Oblique Incidence Response Analysis} 
\begin{figure}[!t]
    \hspace*{0.55cm} 
    \begin{overpic}[width=0.45\textwidth, height=0.15\textwidth, grid=false, tics=5, trim={0cm 0cm 0cm 0cm}, clip]{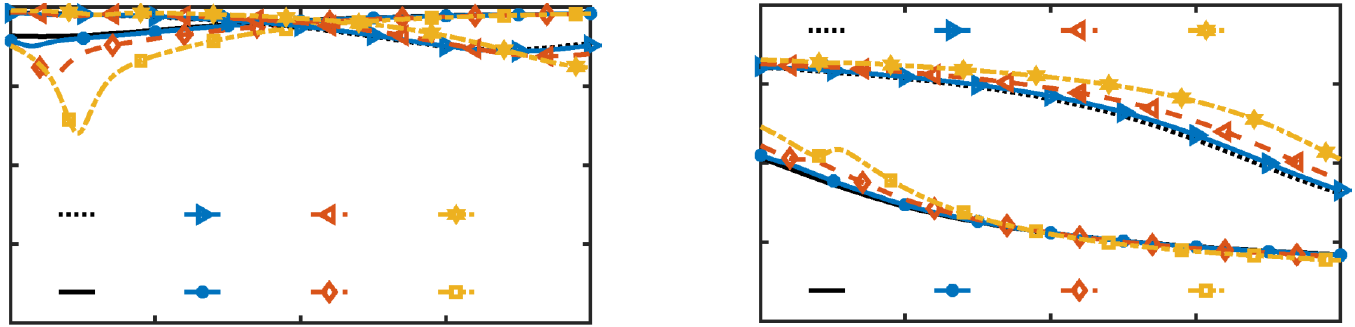}
		\put(-1.15, -2.5){\scriptsize 4.8}
		\put(9.75, -2.5){\scriptsize 5.2}
		\put(20.5, -2.5){\scriptsize 5.6}
		\put(31, -2.5){\scriptsize 6.0}
		\put(41.75, -2.5){\scriptsize 6.4}
		\put(-3.75, 0){\scriptsize -20}
		\put(-3.75, 7.5){\scriptsize -15}
		\put(-3.75, 15.65){\scriptsize -10}
		\put(-3.75, 23.8){\scriptsize -05}
		\put(-3, 32){\scriptsize 00}
		\put(11.75, -6){\scriptsize Frequency (GHz)}
		\put(20.5, -10){\scriptsize (a)}
		\put(-8, 12){\rotatebox{90}{\scriptsize $|\Gamma|$ (dB)}}
		\put(7.0, 2.9){\tiny $00\degree_{\text{off}}$}
		\put(16.5, 2.9){\tiny $15\degree_{\text{off}}$}
		\put(25.5, 2.9){\tiny $30\degree_{\text{off}}$}
		\put(35.5, 2.9){\tiny $45\degree_{\text{off}}$}
		\put(7.0, 11){\tiny $00\degree_{\text{on}}$}
		\put(16.5, 11){\tiny $15\degree_{\text{on}}$}
		\put(25.5, 11){\tiny $30\degree_{\text{on}}$}
		\put(35.5, 11){\tiny $45\degree_{\text{on}}$}
		\put(54.15, -2.5){\scriptsize 4.8}
		\put(65, -2.5){\scriptsize 5.2}
		\put(75.65, -2.5){\scriptsize 5.6}
		\put(86.5, -2.5){\scriptsize 6.0}
		\put(96.25, -2.5){\scriptsize 6.4}
		\put(50.25, 0){\scriptsize -240}
		\put(50.25, 7.5){\scriptsize -120}
		\put(52.75, 15.65){\scriptsize 00}
		\put(51.25, 23.8){\scriptsize 120}
		\put(51.25, 32){\scriptsize 240}
		\put(47, 11.25){\rotatebox{90}{\scriptsize $\angle \Gamma (\text{deg})$}}
		\put(66.75, -6){\scriptsize Frequency (GHz)}
		\put(75.5, -10){\scriptsize (b)}
		\put(62.45, 3){\tiny $00\degree_{\text{off}}$}
		\put(71.65, 3){\tiny $15\degree_{\text{off}}$}
		\put(81.15, 3){\tiny $30\degree_{\text{off}}$}
		\put(90.5, 3){\tiny $45\degree_{\text{off}}$}
		\put(62.45, 29.65){\tiny $00\degree_{\text{on}}$}
		\put(71.65, 29.65){\tiny $15\degree_{\text{on}}$}
		\put(81.15, 29.65){\tiny $30\degree_{\text{on}}$}
		\put(90.5, 29.65){\tiny $45\degree_{\text{on}}$}
    \end{overpic}
    \vspace{14pt}
    \caption{Simulated (a) magnitude and (b) phase of reflection coefficient of the unit cell under normal and oblique y- (TE) polarized plane wave incidences, respectively.}
    \label{Fig. mag phase}
\end{figure}

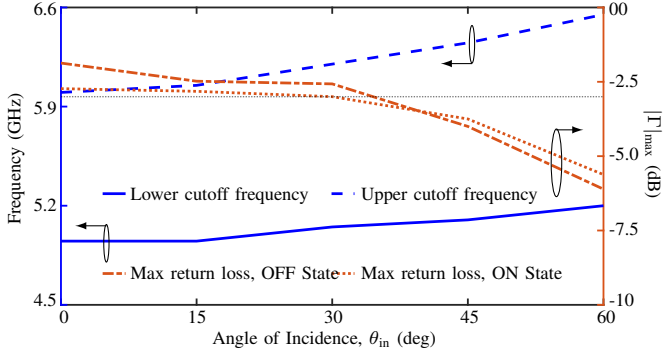
\begin{figure}[!t]
	\centering
	\begin{overpic}[width=0.4\textwidth, grid=false, tics=5, trim={0cm 0cm 0cm 0cm}, clip]{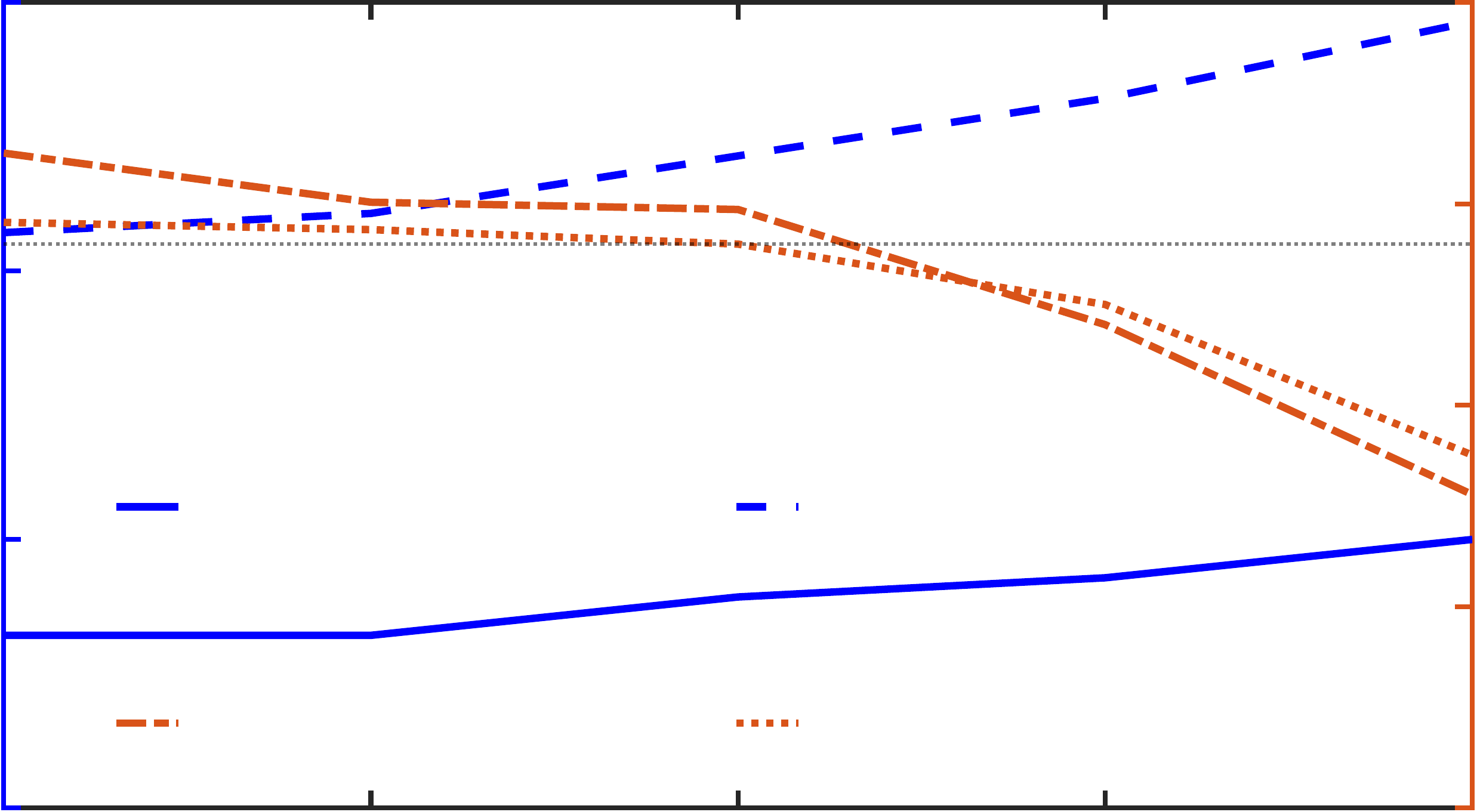}
		\put(90, 20){\tikz{\draw[black, thin] (0,0) ellipse [x radius=0.07, y radius=0.45];}}
		\put(91,32.35){\textcolor{black}{\rotatebox{00}{\vector(1,0){05}}}}
        \put(75, 44.5){\tikz{\draw[black, thin] (0,0) ellipse [x radius=0.035, y radius=0.25];}}
		\put(69.85,44.5){\textcolor{black}{\rotatebox{180}{\vector(1,0){05.75}}}}
        \put(8, 8){\tikz{\draw[black, thin] (0,0) ellipse [x radius=0.035, y radius=0.25];}}
		\put(2.85,14.75){\textcolor{black}{\rotatebox{180}{\vector(1,0){05.75}}}}
		\put(13,19.5){\textcolor{black}{\scriptsize Lower cutoff frequency}}
		\put(55.25,19.5){\textcolor{black}{\scriptsize Upper cutoff frequency}}
		\put(13,5){\textcolor{black}{\scriptsize Max return loss, OFF State}}
		\put(55.25,5){\textcolor{black}{\scriptsize Max return loss, ON State}}
		\put(-4.75,17.5){\textcolor{black}{\scriptsize 5.2}}
		\put(-4.75,35.25){\textcolor{black}{\scriptsize 5.9}}
		\put(-4.75,53.5){\textcolor{black}{\scriptsize 6.6}}
		\put(-4.75,0){\textcolor{black}{\scriptsize 4.5}}
		\put(-9.5, 15){\rotatebox{90}{\scriptsize Frequency (GHz)}}
        \put(100.75,13){\textcolor{black}{\scriptsize -7.5}}
		\put(100.75,26.75){\textcolor{black}{\scriptsize -5.0}}
		\put(100.75,40.25){\textcolor{black}{\scriptsize -2.5}}
		\put(100.75,53.5){\textcolor{black}{\scriptsize 00}}
		\put(100.75,0){\textcolor{black}{\scriptsize -10}}
		\put(107, 36.25){\rotatebox{-90}{\scriptsize $|\Gamma|_{\text{max}}$ (dB)}}
		\put(0,-3){\textcolor{black}{\scriptsize 0}}
		\put(23.5,-3){\textcolor{black}{\scriptsize 15}}
		\put(48.5,-3){\textcolor{black}{\scriptsize 30}}
		\put(73.5,-3){\textcolor{black}{\scriptsize 45}}
		\put(98.5,-3){\textcolor{black}{\scriptsize 60}}
		\put(28,-7){\textcolor{black}{\scriptsize Angle of Incidence, $\theta_{\text{in}}$ (deg)}}
	\end{overpic}
	\vspace{20pt}
	\caption{Oblique incidence stability analysis of the lower and upper cut-off frequencies for 180$\degree \pm$30$\degree$ phase difference between ON and OFF states and maximum reflection loss, respectively.}
	\label{Fig. angular stability}
\end{figure}

The magnitude and phase response for the normal and oblique incidences of the TE polarized wave are shown in Fig.~\ref{Fig. mag phase}. To analyze the stability of the unit cell response, the lower and upper cut-off frequencies for 180$\degree \pm$30$\degree$ phase difference between the ON and OFF states, and the maximum reflection loss with respect to the incidence angles are plotted in Fig.~\ref{Fig. angular stability}. It can be observed that the reflection loss in the ON state increases at the resonance for increase in angle of incidence. However, the response of the unit cell is quite stable for oblique incidence up to $45^{0}$. It can provide a bandwidth of 4.85 to 6.05 GHz with reflection loss less than 3 dB and phase difference of 180$\degree \pm$50$\degree$ for oblique incidence up to $45^{0}$. 

\section{Array Construction and Evaluation of Radiation Patterns}

To show the beam scanning capability of this metasurface, an array of 16 × 10 unit cells has been developed, incorporating a biasing arrangement for every individual element. The radiation patterns are first evaluated theoretically and then validated using full-wave simulations and measurements. 

\subsection{Radiation Pattern Evaluation Using Array Factor and Full-Wave Simulations}
To achieve beamforming from the array with incidence angle $\theta_{\text{in}}$ and reflection angle $\theta_{\text{out}}$ along $\phi=0^{\circ}$ (xz-) plane, i.e., $\phi_{\text{in}}= \phi_{\text{out}} =0^{\circ}$, the generalized Snell’s law can be used to calculate phase gradient $\varphi_{\text{mn}}$ as

    \begin{equation}
		\varphi_{\text{mn}} ={k}_{0}\left\{\hat{\textbf{r}}_{\text{in}}.\vec{\textbf{r}}_{\text{mn}}-\hat{\textbf{r}}_{\text{out}}.\vec{\textbf{r}}_{\text{mn}}\right\},
        \label{eq. 1}
	\end{equation}

\noindent where, $k_0$ is the wavenumber in free space. The parameters $\vec{\textbf{r}}_{\text{mn}}$ is position vector of the $(m,n)th$ element, and $\hat{\textbf{r}}_{\text{in}}$ and $\hat{\textbf{r}}_{\text{out}}$ are the unit vectors from the $(m,n)th$ element to the direction of the source and desired reflection, respectively. 

Using vector algebra
	\begin{equation}\label{eq. 2}
		\hat{\textbf{r}}_{\text{in}} \cdot \vec{\textbf{r}}_{\text{mn}}= p\sin \theta_{\text{in}} ((m-1) \cos \phi_{\text{in}} + (n-1) \sin \phi_{\text{in}}),
	\end{equation}
	and
	\begin{equation}\label{eq. 3}
		\hat{\textbf{r}}_{\text{out}} \cdot \vec{\textbf{r}}_{\text{mn}}= p\sin \theta_{\text{out}} ((m-1) \cos \phi_{\text{out}} + (n-1) \sin \phi_{\text{out}}).
	\end{equation} Subsequently, the calculated phase gradient can be quantized for a 1-bit metasurface adhering to the following conditions

\begin{equation}
\varphi_{1\text{-bit}} = \begin{cases}
0^\circ, & \varphi_{\text{mn}} \notin [90^\circ, 270^\circ) \\
180^\circ, & \varphi_{\text{mn}} \in [90^\circ, 270^\circ)
\end{cases}
\label{eq. 4}
\end{equation}

The far-field radiation patterns for a coding metasurface of size M × N unit cells with periodicity $p$ and above incidence and reflection angles can be calculated using the array factor formula

	\begin{align}
    F_{\text{Total}}(\theta, \phi) 
    &= \sum_{m=1}^{M} \sum_{n=1}^{N} |\Gamma_{\text{1-bit}}| e^{j\varphi_\text{1-bit}} \nonumber \\
    & \times e^{-j k_0 (\hat{\textbf{r}}_{\text{in}}.\vec{\textbf{r}}_{\text{mn}}-\hat{\textbf{r}} \cdot \vec{\textbf{r}}_{\text{mn}})}.
    \label{eq. 5}
    \end{align}

\begin{figure}
    \centering
{\includegraphics[width=0.94\columnwidth, height=0.6\columnwidth]{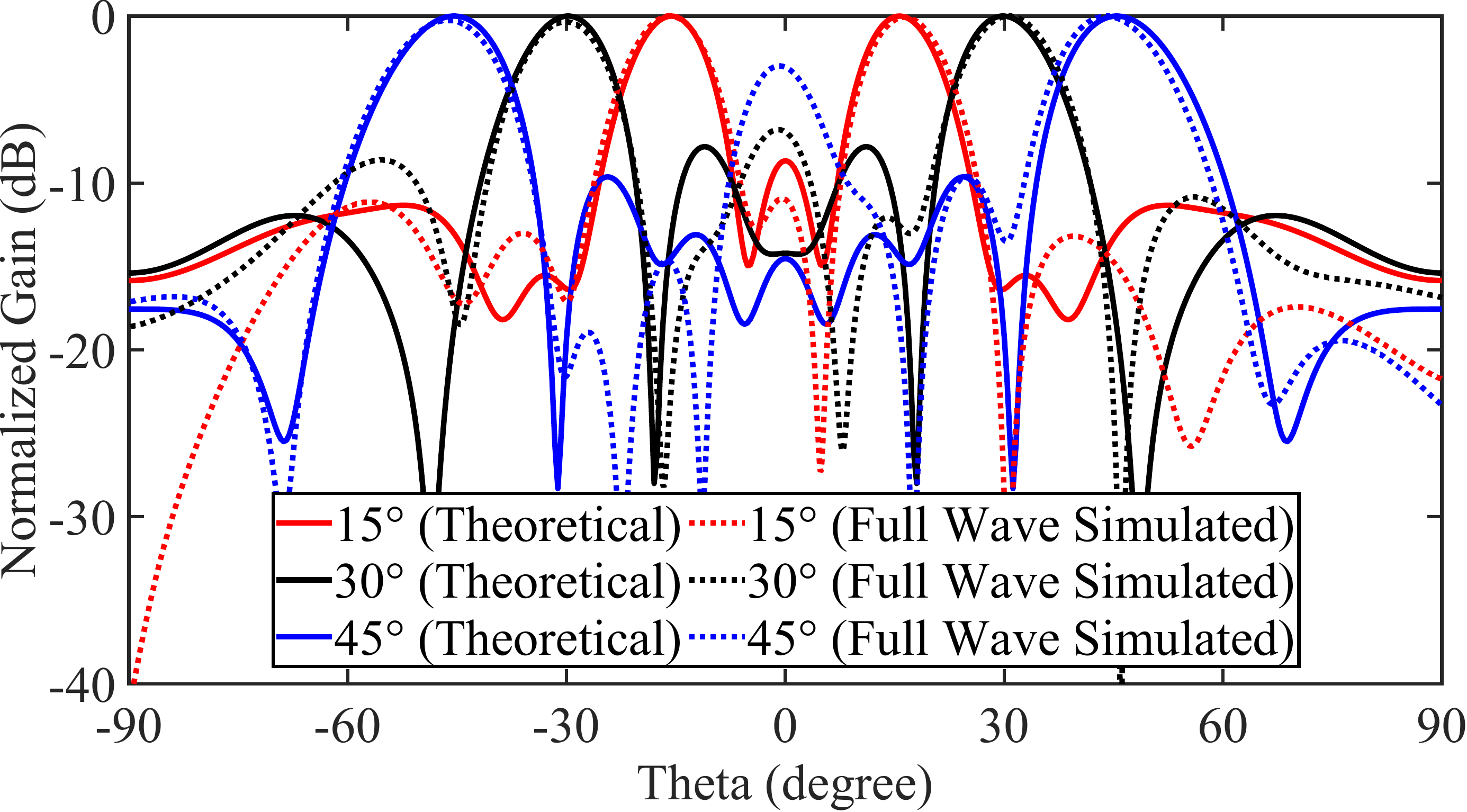}}
 \caption{Array factor-based theoretically calculated radiation patterns compared with full-wave simulated results for normal plane wave incidence at 5.5 GHz.}
    \label{Fig. pattern full wave}
\end{figure}

\noindent where $|\Gamma_{\text{1-bit}}|$ represents the magnitude of the reflection coefficient of 1-bit coding metasurface and $\hat{\textbf{r}}$ is the observation vector. The radiation patterns for the normal plane wave incidence, calculated using (\ref{eq. 5}), are compared with full-wave simulated radiation patterns, obtained from CST MWS. It can be observed from Fig.~\ref{Fig. pattern full wave} that the full-wave simulated results agree well with the theoretically calculated radiation patterns for all reflection angles. However, at specular reflection angle, the full wave results show slight discrepancies from array factor-based calculation. It can be attributed to the mutual coupling of the heterogeneous elements and edge effects due to the finite size of the RIS \cite{10666074}.  

\begin{figure}[!ht]
	\centering
	\begin{overpic}[width=0.4\textwidth,  grid=false, tics=5, trim={0cm 0cm 0cm 0cm}, clip]{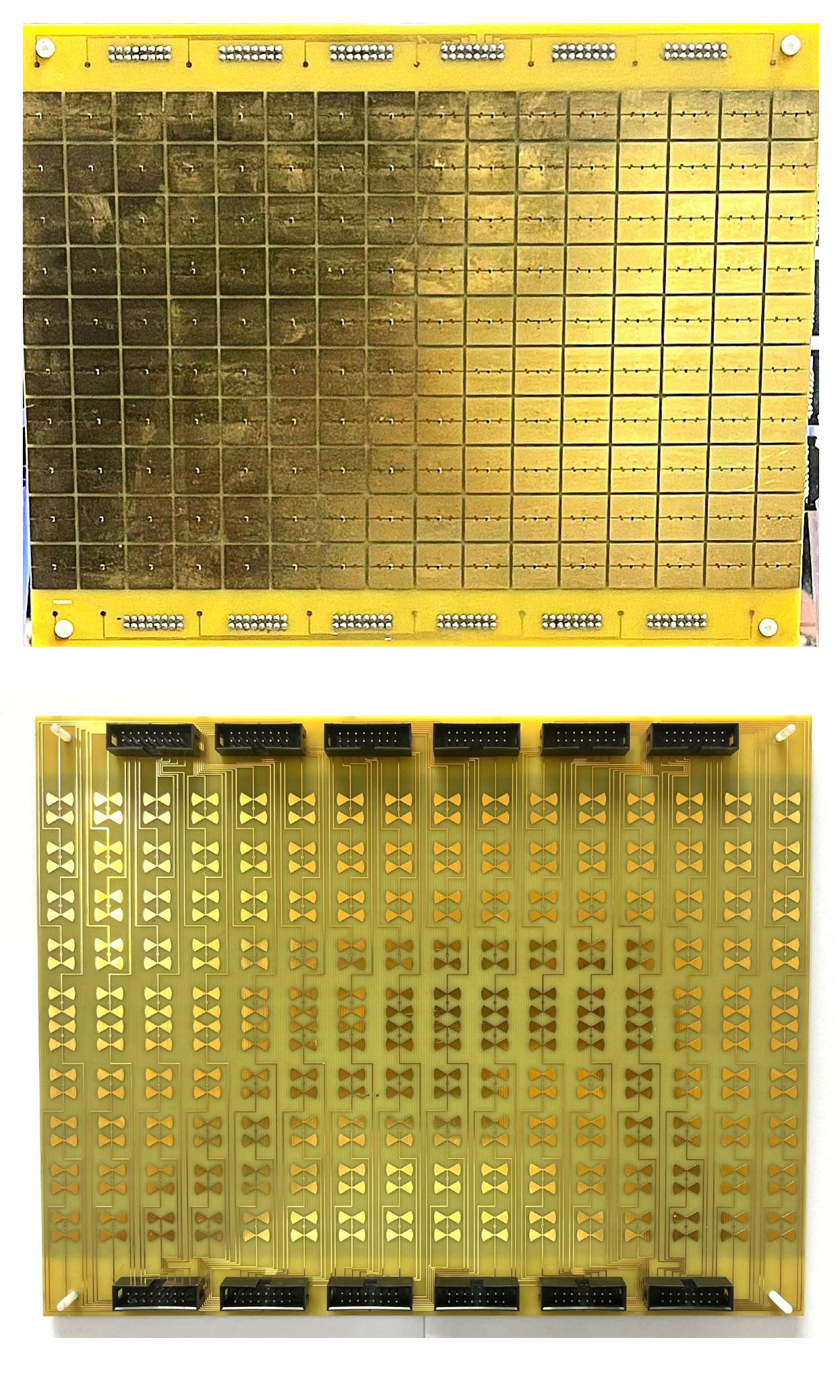}
    \put(29, 51){\scriptsize (a)}
     \put(29, 1){\scriptsize (b)}
	\end{overpic}
	\vspace{3pt}
	\caption{Photograph of the (a) top and (b) backside of the fabricated prototype of the 1-bit coding metasurface-based RIS.}
    \label{Fig. fabricated prototype}
\end{figure}

\begin{figure}[!ht]
	\centering
	\begin{overpic}[width=0.4\textwidth,  grid=false, tics=5, trim={0cm 0cm 0cm 0cm}, clip]{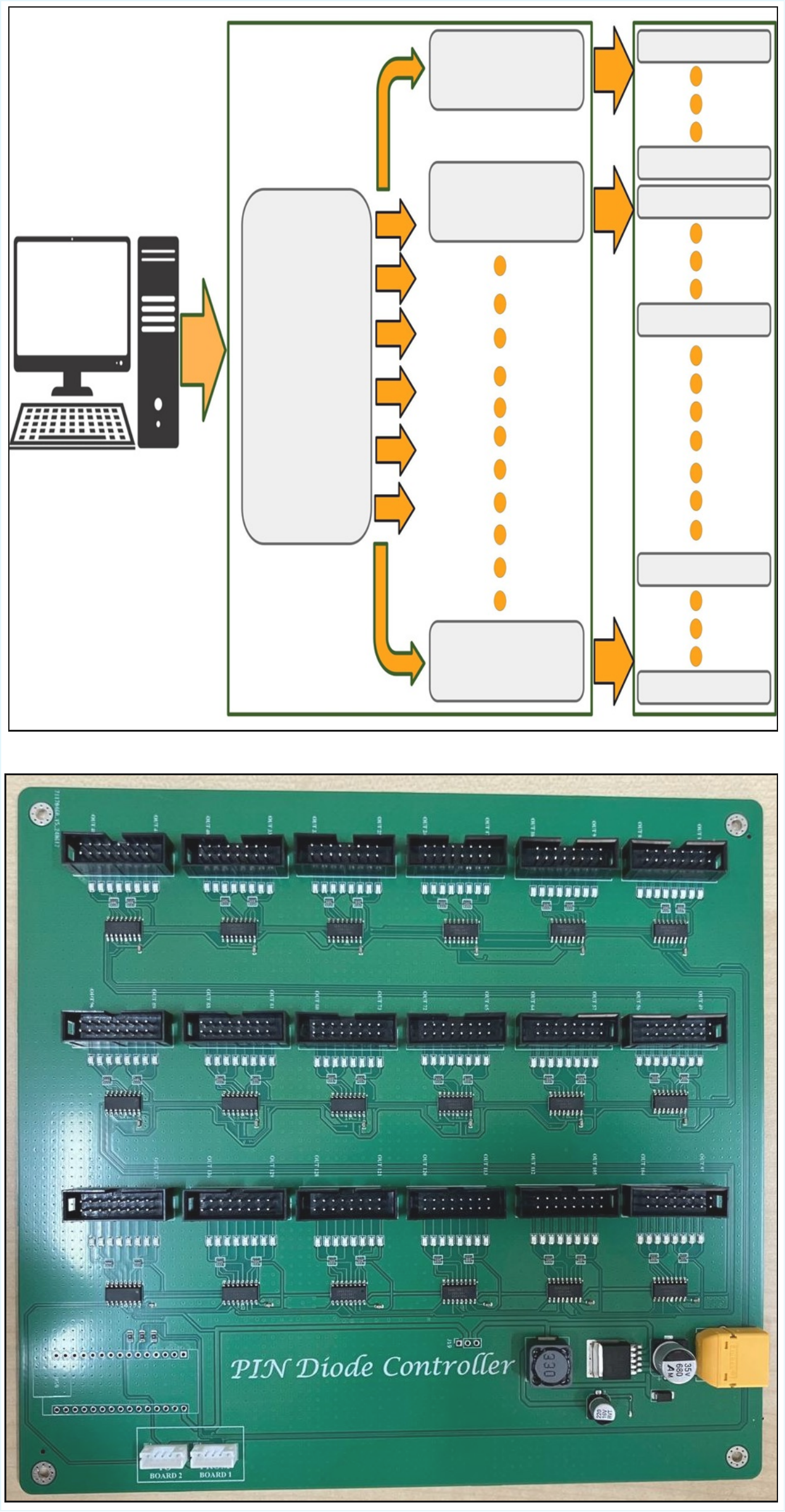}
     \put(2,66){\scriptsize Host PC with}
     \put(1,64){\scriptsize Matlab, Arduino}
     \put(3,62){\scriptsize IDE, and}
     \put(3,60){\scriptsize LabVIEW}
    \put(3.5,81){\tiny Phase}
    \put(3,79){\tiny Pattern}  
    \put(2,77){\tiny Generation}
    \put(16.5,77){\small Arduino} 
    \put(18,75){\small Nano }
    \put(29,96){\footnotesize 74HC595D  } 
    \put(29,87){\footnotesize 74HC595D } 
    \put(29,55){\footnotesize 74HC595D } 
    \put(37,99){\scriptsize Bias lines}
    \put(43,96.5){\tiny $1^{st}$pin diode }
    \put(43,88.5){\tiny $8^{th}$pin diode } 
    \put(43,86){\tiny $9^{th}$pin diode }
    \put(42.25,78.5){\tiny $16^{th}$pin diode} 
    \put(42.15,62){\tiny $153^{rd}$pindiode}
    \put(42.15,54){\tiny $160^{th}$pindiode} 
    \put(26, 50){\scriptsize (a)}
     \put(26, -1.25){\scriptsize (b)}
	\end{overpic}
	\vspace{3pt}
    \caption{a) Block diagram representation of phase change mechanism via control board b) Photograph of the fabricated control board}
    \label{Fig. control circuit}
\end{figure}
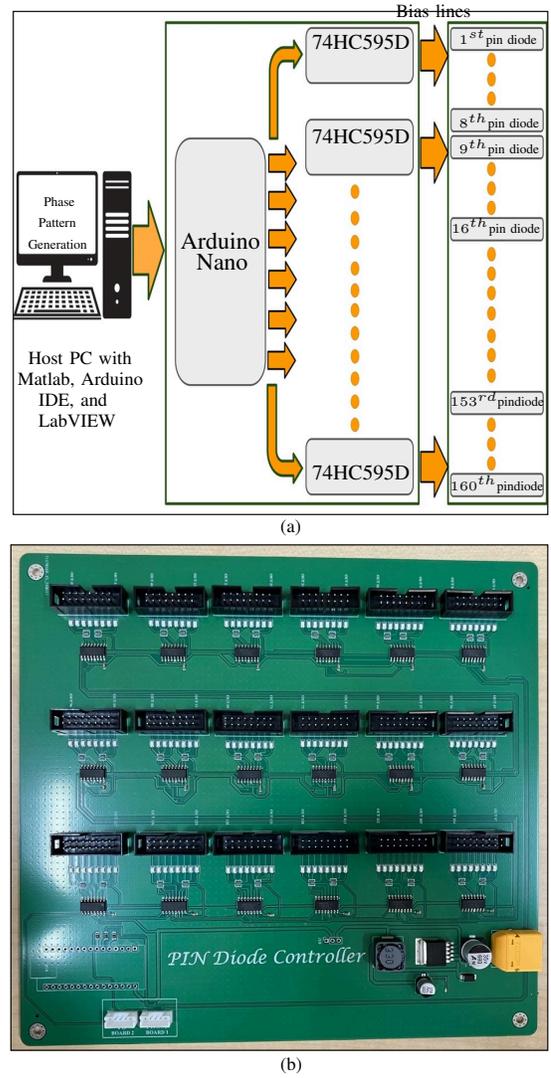

\subsection{Fabrication and Radiation Pattern Measurement} 
The RIS with 16 × 10 unit cells is manufactured using the standard PCB fabrication technique. The top and backside of the fabricated RIS prototype are shown in Fig.~\ref{Fig. fabricated prototype}. A total 160 SMP 1340-040LF PIN diodes are soldered on the top layer of the fabricated RIS. The block diagram of the dynamic phase change mechanism using the Arduino-based control circuit board is shown in Fig.~\ref{Fig. control circuit}(a). It consists of a host PC with LabVIEW and MATLAB API. MATLAB is used to generate the phase pattern, program the Arduino on the control board, and LabVIEW is used to display the LTE application framework's transceiver data (detailed in Section - IV). The phase pattern data calculated from MATLAB are fed to the control board. Fig.~\ref{Fig. control circuit}(b) shows the fabricated control board. It consists of an Arduino Nano microcontroller connected to 20 serial in parallel out (SIPO) 74HC595D shift register ICs (integrated circuits), 160 SMD LEDs, and 160 SMD resistors to control the 160 PIN diodes of the RIS in real time. Each 8-bit shift register can convert serial data to parallel data, which are then fed to 8 PIN diodes. To apply a forward bias voltage of 0.9 V to each PIN diode from a 3.15 V source voltage of a shift register output pin (considering that the Arduino output voltage is 3.3 V and $<$ 0.15 V drop in the shift register from data sheet), the value of each SMD resistor is selected as $\approx$ 56 $\Omega$ using (\ref{eq. 6}).

\begin{equation}
    R = \frac{V_{\text{source}} - V_{\text{LED}} - V_{\text{PIN}}}{I_{F}}
    \label{eq. 6}
\end{equation}

\noindent where $V_{\text{LED}} = 1.8$ V for the SMD LED used in the circuit and $I_F \approx \mathbf{8 \, \text{mA}}$ for the PIN diode operated in forward bias. 

\begin{figure}[!t]
	\centering
	\begin{overpic}[width=0.45\textwidth,  grid=false, tics=5, trim={0cm 0cm 0cm 0cm}, clip]{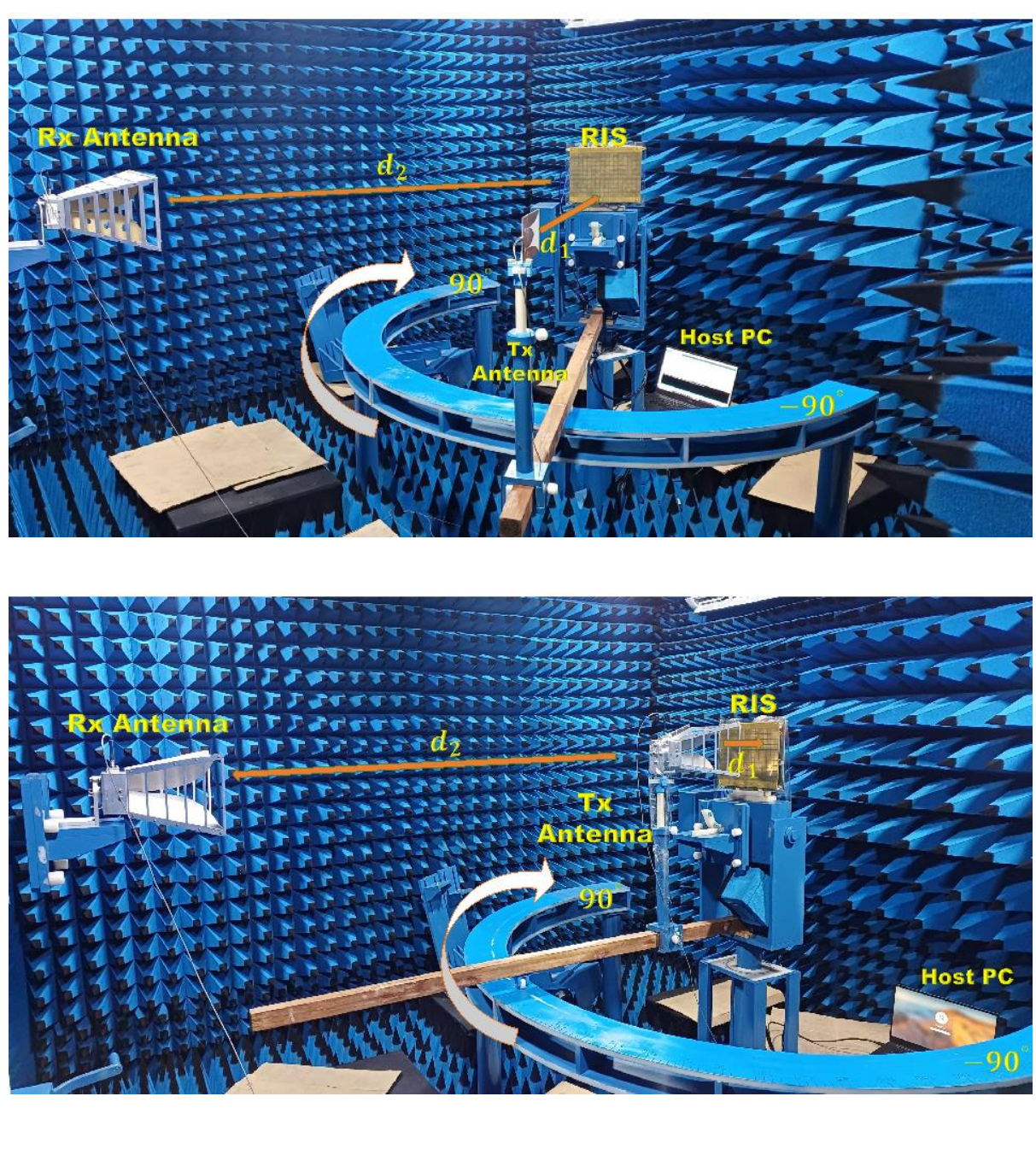}
    \put(45, 50){\scriptsize (a)}
     \put(45, 1){\scriptsize (b)}
	\end{overpic}
	\vspace{3pt}
	\caption{Experimental setup inside the anechoic chamber to characterize the radiation pattern of the RIS for (a) far-field incidence and (b) near-field incidence, respectively. The RIS to receiver antenna distance ($d_{2}$) is 2.2 m. The Transmitter to RIS distance ($d_{1}$) is 2 m and 30 cm for far field and near-field incidences, respectively.}
    \label{Fig. anechoic chamber}
\end{figure}

After successful theoretical calculations and full-wave simulations in CST MWS for beam steering under plane-wave incidence, the fabricated RIS is tested in the anechoic chamber. This time, beam steering for a near-field source from the same RIS is also investigated theoretically and experimentally as a special case, since we will use this configuration for NLOS UE localization and uplink communication in Section IV. Since the fabricated RIS is a small laboratory prototype, one of the antennas is kept at near field to achieve a satisfactory received signal strength by avoiding the extra path loss and loss due to the occurrence of grating lobes. The experimental setups for radiation pattern measurement under both the plane wave and near-field source incidences inside the anechoic chamber are shown in Fig.~\ref{Fig. anechoic chamber}. Both the RIS and the receiving antenna are mounted on a rotating platform with the antenna directed toward the RIS. The transmitting antenna is kept stationary at $\theta = 0^{\circ}$, but at 2 m ($\approx2D^2/\lambda_{0}$) and 0.3 m for plane wave and near-field incidence, respectively. The near-field distance is selected using the same method explained in \cite{kundu2022single}. The radiation pattern for near-field incidence can be calculated using \cite{nayeri2018reflectarray}

\begin{equation}\label{fifth}
		\begin{split}
			F_\text{Total}(\theta, \phi) &= \sum_{m=1}^{M} \sum_{n=1}^{N} (\cos{\theta})^{q_e}\frac{\big[\cos{\theta_f(mn)}\big]^{q_f}}{r_{\text{mn}}^{t}} |\Gamma_{\text{1-bit}}| e^{j\varphi_\text{1-bit}} \\
            &{}\times \big[\cos{\theta_e(mn)}\big]^{q_e}e^{-j k_0 (|\text{r}_\text{{mn}}^{t}|-\vec{\textbf{r}}_{\text{mn}} \cdot \hat{\textbf{r}})},
		\end{split}
	\end{equation}

\noindent where $\theta_f(mn)$ represents the angular offset of the feed horn relative to each metasurface element
	\begin{equation}\label{sixth}
		\theta_f(mn) = \tan^{-1} \left( \frac{\sqrt{(x_f - x_m)^2 + (y_f - y_n)^2}}{z_f} \right).
	\end{equation}

$r_{\text{mn}}^{t}$ is the Euclidean distance from the phase centre ($x_f, y_f, z_f$) of the near-field source to the $(m, n)th$ element on the RIS aperture expressed as    
	\begin{equation}\label{second}
		r_{\text{mn}}^{t} = |\vec{\textbf{r}}_{\text{f}}-\vec{\textbf{r}}_{\text{mn}}| = \left\{ \sqrt{(x_f - x_m)^2 + (y_f - y_n)^2 + z_f^2} \right\}.
	\end{equation}

    The feed's angular illumination profile is modelled using a cosine power function,$\big[\cos{\theta_f(mn)}\big]^{q_f}$. By fitting to the measured radiation pattern of our feed horn, the value of $q_f$ is 7.

\begin{figure}[!t]
    \hspace*{1cm} 
    \begin{overpic}[width=0.4\textwidth, height=0.6\textwidth, grid=false, tics=5, trim={0cm 0cm 0cm 0cm}, clip]{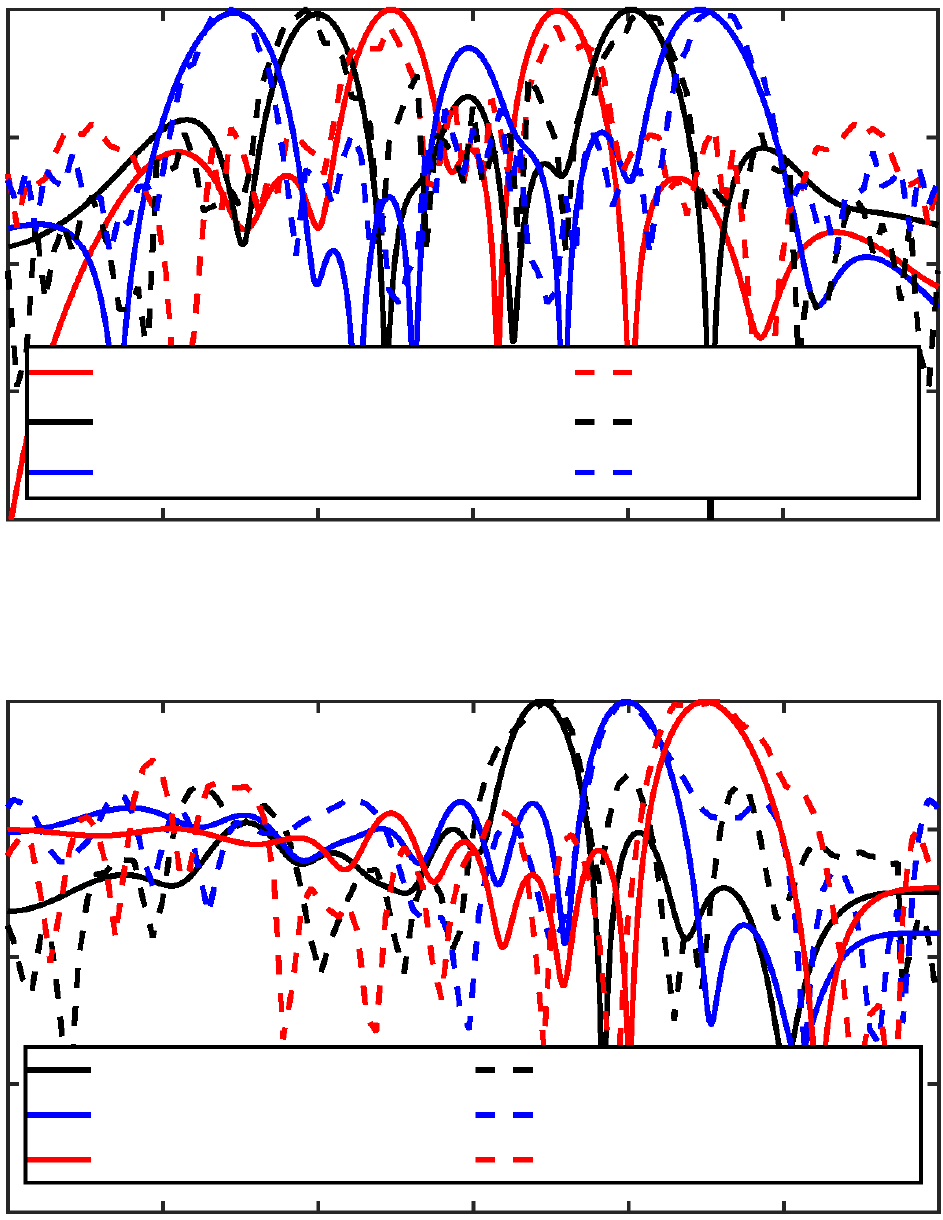}
		\put(-1.5,55){\scriptsize -90}
		\put(9.65,55){\scriptsize -60}
		\put(20.5,55){\scriptsize -30}
		\put(32.25,55){\scriptsize 00}
		\put(43.25,55){\scriptsize 30}
		\put(54.2,55){\scriptsize 60}
		\put(65.2,55){\scriptsize 90}
		\put(-1.5,-1.8){\scriptsize -90}
		\put(9.65,-1.8){\scriptsize -60}
		\put(20.5,-1.8){\scriptsize -30}
		\put(32.25,-1.8){\scriptsize 00}
		\put(43.25,-1.8){\scriptsize 30}
		\put(54.2,-1.8){\scriptsize 60}
		\put(65.2,-1.8){\scriptsize 90}	
		\put(-3,0){\scriptsize -40}
        \put(-3,10.25){\scriptsize -30}
        \put(-3,20.8){\scriptsize -20}
        \put(-3,31.25){\scriptsize -10}
        \put(-2.25,41){\scriptsize 00}
        \put(-3,57){\scriptsize -40}
        \put(-3,67.25){\scriptsize -30}
        \put(-3,77.8){\scriptsize -20}
        \put(-3,88.25){\scriptsize -10}
        \put(-2.25,98){\scriptsize 00}
        \put(28,-6){\scriptsize Theta (deg)}
        \put(32.2,-10){\scriptsize (b)}
        \put(28,51){\scriptsize Theta (deg)}
        \put(32.2,46.5){\scriptsize (a)}
		\put(-7.5, 11){\rotatebox{90}{\scriptsize Normalized Gain (dB)}}
		\put(-7.5, 68){\rotatebox{90}{\scriptsize Normalized Gain (dB)}}
        \put(7,11.6){\scriptsize $15\degree$ (Actual Phase A.F.)}
        \put(7,7.75){\scriptsize $30\degree$ (Actual Phase A.F.)}
        \put(7,4.1){\scriptsize $45\degree$ (Actual Phase A.F.)}
        \put(39,11.6){\scriptsize $15\degree$ (Measured)}
        \put(39,7.75){\scriptsize $30\degree$ (Measured)}
        \put(39,4.1){\scriptsize $45\degree$ (Measured)}  
        \put(7,68.6){\scriptsize $15\degree$ (Full Wave Simulated)}
        \put(7,64.75){\scriptsize $30\degree$ (Full Wave Simulated)}
        \put(7,60.75){\scriptsize $45\degree$ (Full Wave Simulated)}
        \put(46,68.6){\scriptsize $15\degree$ (Measured)}
        \put(46,64.75){\scriptsize $30\degree$ (Measured)}
        \put(46,60.75){\scriptsize $45\degree$ (Measured)}    
    \end{overpic}
    \vspace{32pt}
    \caption{Radiation patterns of the RIS for different reflection angles. (a) Comparison of full-wave simulated and experimentally measured results under normal plane wave incidence. (b) Comparison of array factor-based theoretically calculated and experimentally measured results under normal incidence of near field source.}
    \label{Fig. normalized patterns}
\end{figure}

\begin{figure}[!t]
	\centering
    \begin{overpic}[width=0.48\textwidth, height=0.175\textwidth, grid=false, tics=5, trim={0cm 0cm 0cm 0cm}, clip]{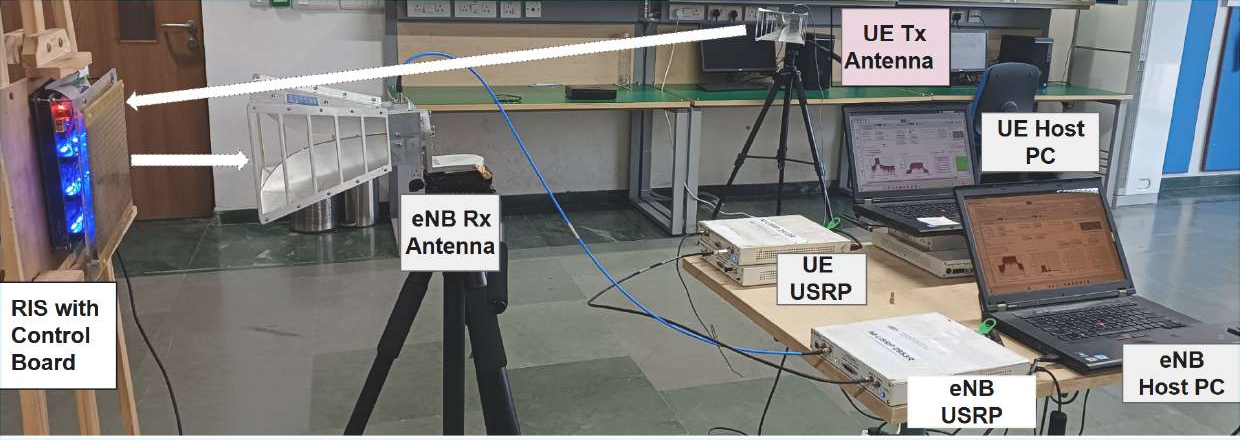}
	\end{overpic}
	\vspace{3pt}
	\caption{ Snap of the RIS-assisted LTE hardware setup used for UE localization and uplink communication performance evaluation.}
    \label{Fig. Indoor}
\end{figure}

\begin{table}[H] 
\centering
\caption{Utilized Hardware Details}
\addtolength{\tabcolsep}{4pt}
\renewcommand{\arraystretch}{1.2}
\fontsize{8pt}{8pt}\selectfont
\begin{tabular}{ |c|c|c| } 
 \hline
 \textbf{Name} &  \textbf{Description} \\
 \hline
Horn Antenna gain at 5.5 GHz & 12dB \\ 
 \hline
Horn Antenna Frequency Range & 0.8-18 GHz \\
 \hline
Antenna Dimension in mm (L$\times$ W $\times$ H) & $228 \times 244 \times 160.5$
 \\
  \hline
Software defined radio & NI-USRP 2953R \\
 \hline
RIS dimension in mm (L$\times$ W $\times$ H) & $256 \times 180 \times 3.2$ \\
 \hline
\end{tabular}
\label{table:1}
\end{table}

The results, presented in Fig.~\ref{Fig. normalized patterns}, show normalized patterns with the peak set at 0 dB. As expected, the main beam is steered at the desired reflected angles $15^{\circ}$, $30^{\circ}$, $45^{\circ}$ when the RIS is configured with their corresponding phase patterns. It can be seen that the measurement results agree well with the full-wave simulation results, affirming the accuracy of the design and its function.

\section{Demonstration of RIS-Assisted UE Localization and Uplink Communication}

In this section, experimental studies are carried out to validate the real-time operation of the fabricated RIS in UE localization and uplink communication. 

\subsection{Experimental Setup and Hardware Specifications} 
Fig.~\ref{Fig. Indoor} presents a proof-of-concept RIS-assisted LTE uplink experimental testbed to evaluate UE localization and communication performance of our RIS in indoor NLOS scenarios. The user equipment (UE) and evolved Node~B (eNB) are implemented using National Instruments$^{\text{TM}}$(NI) USRP-2953R platforms interfaced with host PCs and equipped with high-gain horn antennas mounted on tripods. The key specifications of the utilized hardware components are detailed in Table II. LabVIEW NXG’s LTE Application Framework is employed to implement an LTE physical layer, enabling real-time transmission and reception using USRP-based SDRs. The fabricated RIS with its dedicated control board is placed between the UE and eNB to manipulate the wireless channel. The dual-channel capability of the USRP is exploited by configuring one RF chain for uplink data transmission and the other to establish a dedicated synchronization and control link between the UE and the eNB. Since LTE uplink transmission is inherently eNB-controlled and requires prior downlink signaling for synchronization, system information acquisition, resource allocation, and timing alignment, this secondary link ensures that a standard-compliant uplink operation can be realized. In a deployed 5G NR scenario, this synchronization would be maintained via a separate control channel or a robust low-rate carrier, justifying this testbed architecture. Now, to realize the UE localization in an uplink scenario, the UE transmit antenna is positioned in an NLOS configuration at a distance of $d_2= 5$~m from the RIS, while the eNB receiver antenna is located approximately $d_1= 30$~cm in front of the RIS as shown in Fig.~\ref{Fig. Indoor}. The LabVIEW NXG eNB code is augmented with an additional block to interfacing with an Arduino Nano for real-time RIS phase control. 

\begin{figure}
    \centering
    \begin{overpic}[width=0.94\columnwidth,  grid=false, tics=3, trim={0cm 0cm 0cm 0cm}, clip]{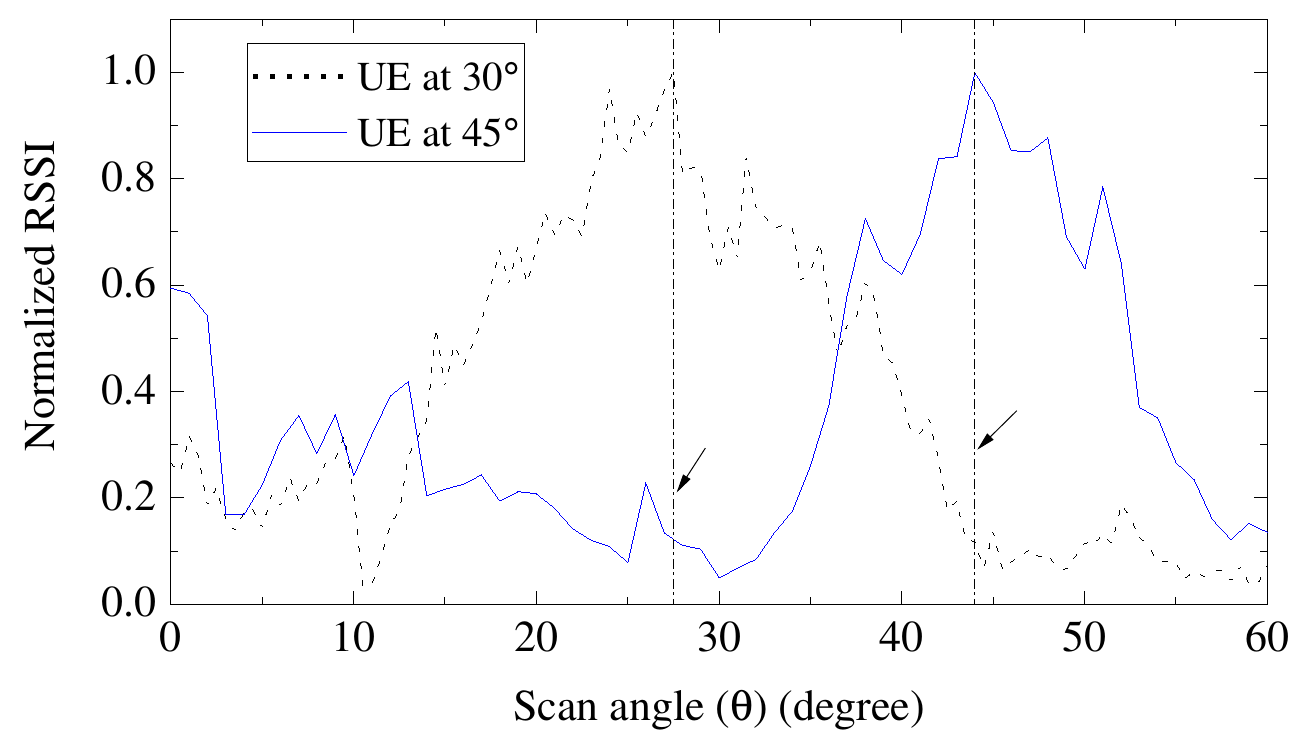}

        \put(52,23.75){\tiny $\theta_{est,1}$}
        \put(76,26.75){\tiny $\theta_{est,2}$}

    \end{overpic}

    \caption{Received signal strength when phase distribution on RIS is configured for beam angle sweep from 0$\degree$ to 60$\degree$ with 1.5$\degree$ steps.}
    \label{Fig. pattern doa}
\end{figure}

\subsection{UE Localization and Beam Alignment Protocol} 

To establish the NLOS uplink, a codebook-based beam alignment protocol is implemented. Since the UE location is initially unknown, the RIS performs a blind angular search to determine the optimal reflection angle ($\theta_{opt}$) that maximizes the link budget. A codebook of quantized phase configurations, $\mathcal{C} = \{\Phi_1, \Phi_2, ..., \Phi_N\}$, is pre-loaded onto the control board, where each vector $\Phi_n$ corresponds to a specific steering angle in $1.5^\circ$ increment within the $0^\circ- 60^\circ$ field of view. During the initialization phase, the UE transmits a continuous uplink reference signal at 5.5 GHz. Simultaneously, the RIS controller sequentially iterates through the phase codebook. The eNB receiver monitors the received signal strength indicator (RSSI) corresponding to each interval. The system estimates the UE's angular position relative to the RIS by identifying the configuration index that maximizes the received power: $\Phi_{opt} = \arg \max_{\Phi \in \mathcal{C}} (\text{RSSI})$, as evidenced by the distinct angular peaks plotted in Fig.~\ref{Fig. pattern doa}. To quantify the precision of the proposed localization protocol, the root mean square error (RMSE) is calculated based on the angular estimation peaks shown in Fig.~\ref{Fig. pattern doa} using 

\begin{equation}
\text{RMSE} = \sqrt{\frac{1}{K} \sum_{i=1}^{K} (\theta_{est,i} - \theta_{true,i})^2}
\label{eq. 7}
\end{equation}

\noindent where $K$ is total number of measurements, $\theta_{true,i}$ and $\theta_{est,i}$ are the true positions of the UE and estimated angles using the above method, respectively. From the results of above experiment for UE positions at $30^{\circ}$ and $45^{\circ}$ the RMSE is obtained as $2.06^{\circ}$. Thus, despite simplicity, this process effectively \textit{localizes} the user in the angular domain, allowing the RIS to latch the optimal phase mask and establish a high-gain NLOS uplink connection. Although advanced direction-of-arrival (DoA) estimation algorithms exist, this deterministic codebook-based method offers low computational overhead while maintaining an acceptable angular precision.

\begin{figure}[!t]
	\centering
	\begin{overpic}[width=0.48\textwidth, height=0.32\textwidth,  grid=false, tics=5, trim={0cm 0cm 0cm 0cm}, clip]{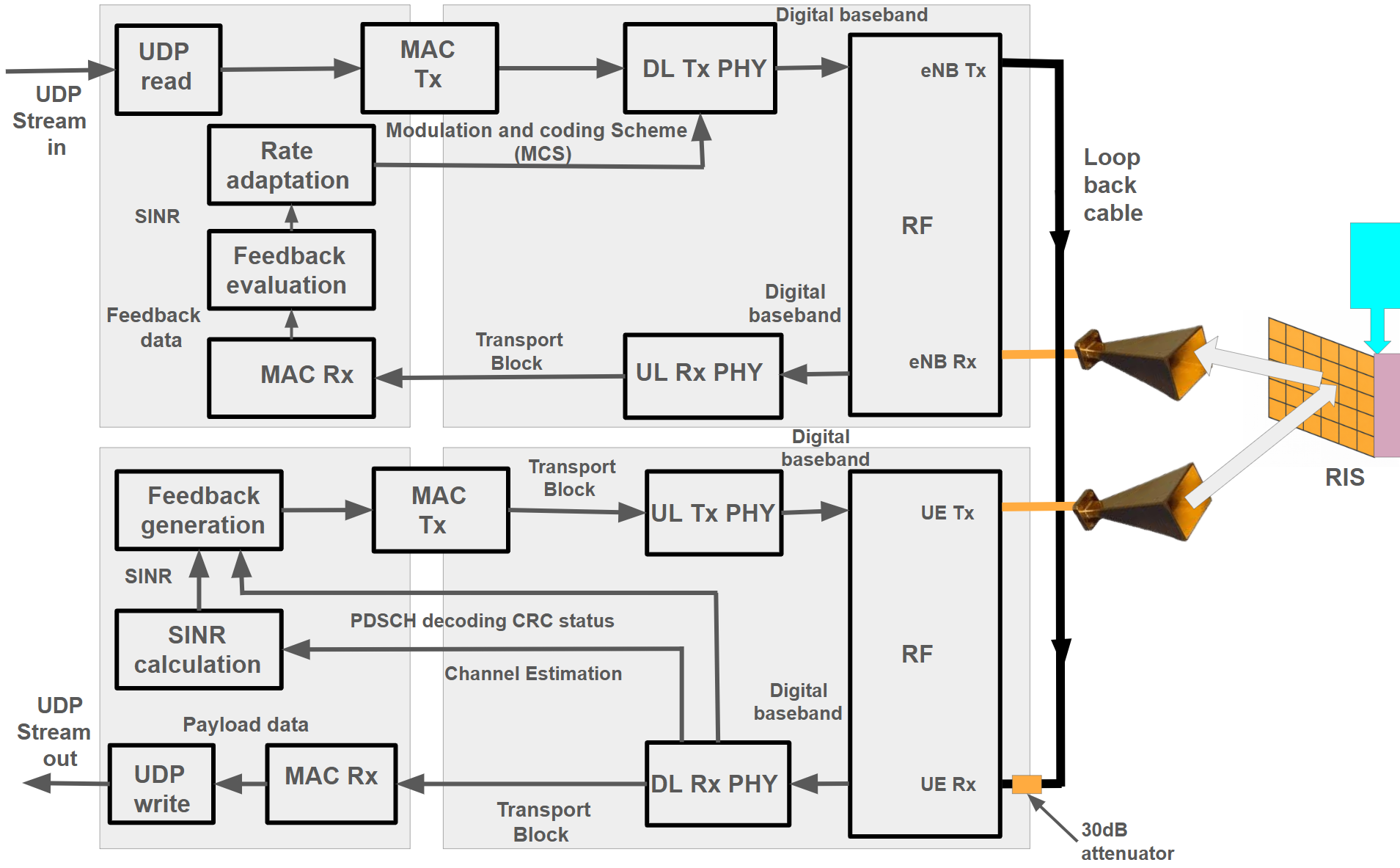}
	\end{overpic}
	\vspace{3pt}
    \caption{Block diagram representation of the RIS-aided wireless system in uplink (UL) operation mode with NI-USRP double-device setup.}
    \label{fig. uplink USRP}
\end{figure}

Following the successful localization and phase mask selection, the system transitions into an active communication mode. Fig.~\ref{fig. uplink USRP} depicts the block diagram and signal flow within the LTE Application Framework module, implemented using two NI USRP-2953R devices for our RIS-aided uplink experimentation. Data transmission is initiated via a UDP stream mapped to the LTE Transport Block (TB). The reconfigurable LTE PHY layer handles OFDM modulation, resource mapping, and channel estimation, generating a 20 MHz digital uplink waveform at 5.5 GHz. This signal is radiated through the USRP RF0 chain toward the RIS. Upon reflection by the RIS, the eNB receiver performs time/frequency synchronization, equalization, and resource element (RE) demapping. Finally, the decoded TB is extracted to close the loop for end-to-end performance analysis. To evaluate link performance, Signal-to-Interference-plus-Noise Ratio (SINR) estimation is carried out using channel estimates derived from cell-specific or UE-specific reference signals. The estimated SINR is utilized for rate adaptation, wherein the modulation and coding scheme (MCS) is dynamically selected to maintain a low block error rate (BLER). In addition, feedback evaluation is performed to extract wideband and subband SINR metrics, along with ACK/NACK indicators, thereby enabling adaptive transmission and improving the reliability of the RIS-assisted uplink communication.

\subsection{Link Performance and Real-Time Communication Analysis} 

To compare the experimental findings with theoretical limits, the analytical received signal power ($P_r$) is first evaluated. As evident from Fig.~\ref{Fig. Indoor}, the RIS operates within the near-field region of eNB antenna. Consequently, to take into account the phase variation of the spherical wave across the RIS surface, we adopt a physics-based summation model \cite{tang2020wireless} that accounts for the individual unit cell distances 

\begin{equation}
\begin{split}
P_r &= P_t \frac{G_t G_r G_{unit}^2 \lambda^2 d_x d_y}{64 \pi^3} \cdot L_{PE} \\
&\quad \cdot \left| \sum_{m=1}^{M} \sum_{n=1}^{N} \frac{\sqrt{F_{m,n}^{combine}}}{r_{mn}^t r_{mn}^r} \exp(j\Psi_{m,n}) \right|^2
\end{split}
\label{eq:received_power}
\end{equation}

\noindent where the system parameters are defined by our experimental setup at 5.5 GHz ($\lambda = 0.0545$ m) as follows. The transmit power is $P_t = -7.87$ dBm applied through USRP, horn antenna gains $G_t = G_r = 12$ dBi (Table II). The physical area of each unit cell provides an effective unit cell gain of $G_{unit} = 4\pi \frac{d_x d_y}{\lambda^2} \approx 0.34$ dBi. The parameters $r_{mn}^t$ and $r_{mn}^r$ represent the specific Euclidean distances from the center of the $(m,n)$-th unit cell to the phase centers of the Tx and Rx antennas, respectively. $r_{mn}^t$ can be calculated using (\ref{second}). $r_{mn}^r$ can be calculated similarly as

\begin{equation}
r_{mn}^r = \sqrt{(x_{r} - x_m)^2 + (y_{r} - y_n)^2 + z_{r}^2}
\label{eq:rx_distance}
\end{equation}

\noindent where $(x_{r}, y_{r}, z_{r})$ represents the spatial coordinate of the receiver, which for a $45^\circ$ steering angle at distance $d_2$ can be evaluated as $(d_2 \sin 45^\circ, 0, d_2 \cos 45^\circ)$. The term $F_{m,n}^{combine}$ denotes the combined normalized power radiation pattern. It takes into account the amplitude taper at the unit cells located at the outer edges of the $16 \times 10$ array compared to the center cells due to the near-field source incidence. It is mathematically formulated as the product of the transmit antenna pattern, the unit cell's effective aperture, and the receive antenna pattern using

\begin{equation}
F_{m,n}^{combine} \approx \cos^{q_t}(\theta_t^{m,n}) \cdot \cos(\theta_{in}^{m,n}) \cdot \cos(\theta_{out}^{m,n}) \cdot \cos^{q_r}(\theta_r^{m,n})
\label{eq:f_combine}
\end{equation}

\noindent where $\theta_{in}^{m,n}$ and $\theta_{out}^{m,n}$ are the angles of incidence and reflection relative to the normal of the $(m,n)$-th unit cell. The angles $\theta_t^{m,n}$ and $\theta_r^{m,n}$ are measured relative to the boresight of the Tx and Rx horn antennas. The directivity exponents are approximated in Section III as $q_t = q_r \approx 7$. The maximum theoretical gain is achieved when the applied phase shifts, $\Psi_{m,n}$, perfectly cancel the cascaded spatial channel phases, effectively reducing the complex exponential term to unity. By iterating the double summation in Equation (\ref{eq:received_power}) - before the magnitude squaring - over all 160 discrete elements for our specific Tx-RIS-Rx arrangement ($d_1 = 0.3$ m, $d_2 = 5$ m steered to $45^{\circ}$), the geometric amplitude accumulation is estimated at approximately 74.19. Due to the near-field power tapering, this value is lower than that of an ideal uniform plane-wave assumption. To account for the discrete phase states of our 1-bit RIS architecture, we incorporate the phase-error loss term, $L_{PE}$, which represents the gain degradation relative to an ideal continuous-phase surface. For a 1-bit quantization scheme, $L_{PE}$ is analytically determined to be approximately -3.92 dB using \cite{jeong2022improved}

\begin{equation}
L_{PE}(\theta_i, \phi_i, \theta_r, \phi_r) = \frac{\left| \sum_{n=1}^{N} e^{j[\varphi_{rn} + \varphi_{en}(\theta_i, \phi_i)]} \right|^2}{\left| \sum_{n=1}^{N} e^{j\varphi_{rn}} \right|^2}
\label{eq:phase_error_loss}
\end{equation} 

\noindent Where $\varphi_{rn}$ and $\varphi_{en}(\theta_i, \phi_i)$ are the required ideal reflection phase for the \textit{n}-th unit cell to steer the beam in the desired direction, and the phase error (i.e., $\varphi_{rn}-\varphi_{quantized}$) of the \textit{n}-th unit cell, respectively. Applying these parameters to the analytical model yields a theoretical maximum received power of $P_r \approx -43.87$ dBm. In a standard 20 MHz LTE channel with a thermal noise floor of approximately -94 dBm, this establishes a theoretical Signal-to-Noise Ratio (SNR) ceiling of nearly 50 dB.

\begin{figure}
    \centering
    \begin{overpic}[width=0.48\textwidth, height=0.27\textwidth, grid=false, tics=5, trim={0cm 0cm 0cm 0cm}, clip]{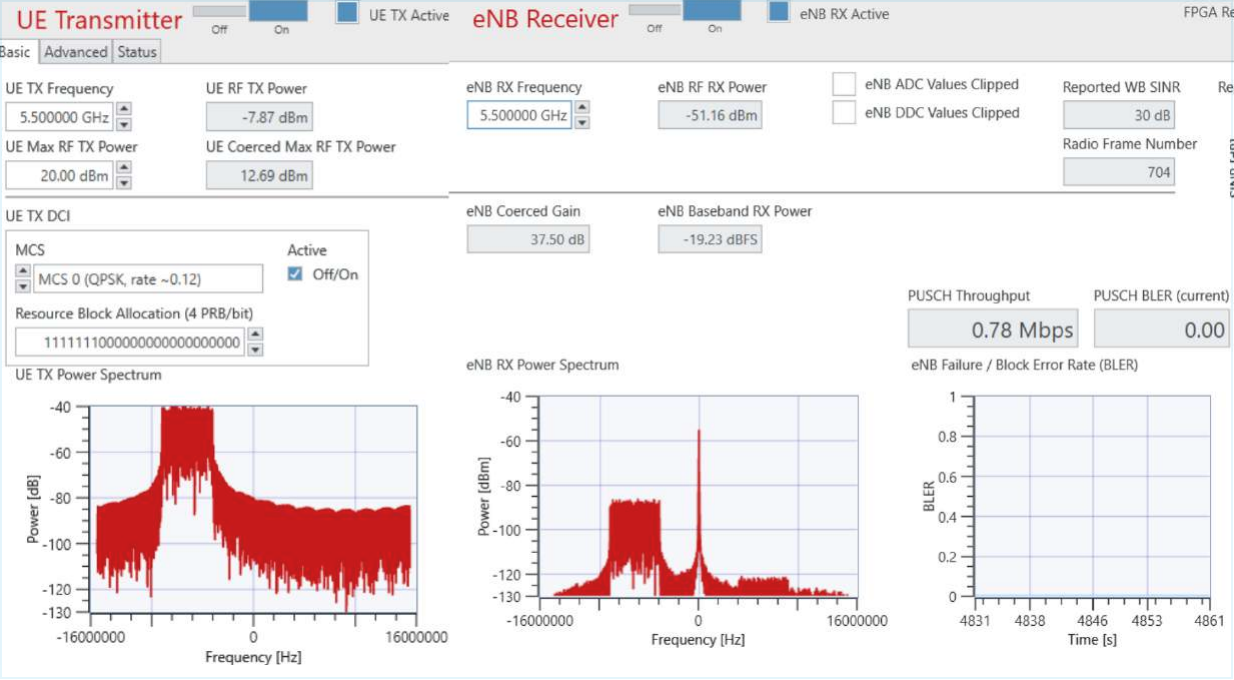}
    \end{overpic}
    \vspace{2pt}
    \caption{Real-time GUI snapshot of the LTE Application Framework showing UE uplink transmission and RIS-assisted eNB reception at 5.5 GHz.}
    \label{fig:spectrum}
\end{figure}

The real-time link performance is experimentally evaluated by first analyzing the analog radio frequency (RF) profile. Fig.~\ref{fig:spectrum}  (lower-middle) shows the uplink received power spectrum at the eNB. It can be observed that the reflected signal rises $\approx 37.5$ dB above the noise floor. Since the spectrum analyzer displays the Power Spectral Density (PSD) rather than the total integrated channel power, the visual plateau and the noise floor rest at approximately -84.5 dBm and -122 dBm per subcarrier, respectively. This yields a 37.5 dB signal lift. For the 20 MHz LTE waveform with approximately 1,200 active subcarriers, the total integrated power scales by a factor of $10\log_{10}(1200) \approx 30.8$ dB. Consequently, the total integrated received signal power and noise floor are evaluated as approximately -53.7 dBm and -91.2 dBm, respectively. This measured total signal power of -53.7 dBm can be directly compared with the theoretical maximum of -43.87 dBm derived using (\ref{eq:received_power}). The discrepancy of roughly 9.83 dB between the theoretical and measured total power can be attributed to the practical hardware implementation losses not taken into account in (\ref{eq:received_power}). Hardware impairment losses include mainly FR4 dielectric dissipation, PIN diode resistance, and coaxial cable attenuation. The loss contribution of FR4 dielectric and PIN diodes is quantified as $\approx$ 3 dB from measurement result by comparing the main lobe maximum of $\theta_{\text{in}}= \theta_{\text{out}} = 0^{\circ}$ case with a perfect electric conductor (PEC) plate of same size. The contribution of cable loss is quantified as 6.87 dB at 5.5 GHz by comparing the measured $|S_{21}|$ (dB) obtained using a VNA calibrated without the utilized cables against a measurement where the utilized cables are added as the device under test (DUT). The remaining $\approx$ 3 dB discrepancy between theoretical and measured SNR can be mainly attributed to a higher measured noise floor in the indoor environment compared to the ideal thermal noise floor, destructive interference due to multipath fading, and system non-idealities, such as USRP RF front-end mismatches. The digital communication quality is further evaluated using the wideband SINR (WB SINR). The LTE Application Framework reports a WB SINR of 30.40 dB (Fig.~\ref{fig:spectrum}, top-right). While the 37.5 dB signal lift represents the analog power gap, the WB SINR is the effective baseband metric evaluated after the USRP decodes the 20 MHz channel. Therefore, it accounts for the remaining real-world channel impairments, including lab interference, residual phase-error distortions, and USRP impairments (such as local oscillator phase noise and quantization errors). During the transmission of a UDP video stream, the system's rate adaptation algorithm selects the most robust modulation and coding scheme (here, MCS 0 (QPSK, coding rate $\approx 0.12$)). While the 30.40 dB WB SINR indicates a high-quality link, the heavy redundancy of the MCS 0 code rate restricts the active payload capacity. As a result, a stable baseband throughput of 0.78 Mbps and a block error rate (BLER) of $< 0.1\%$ are maintained.

\begin{figure}[!t]
	\centering
	\begin{overpic}[width=0.48\textwidth, height=0.28\textwidth, grid=false, tics=5, trim={0cm 0cm 0cm 0cm}, clip]{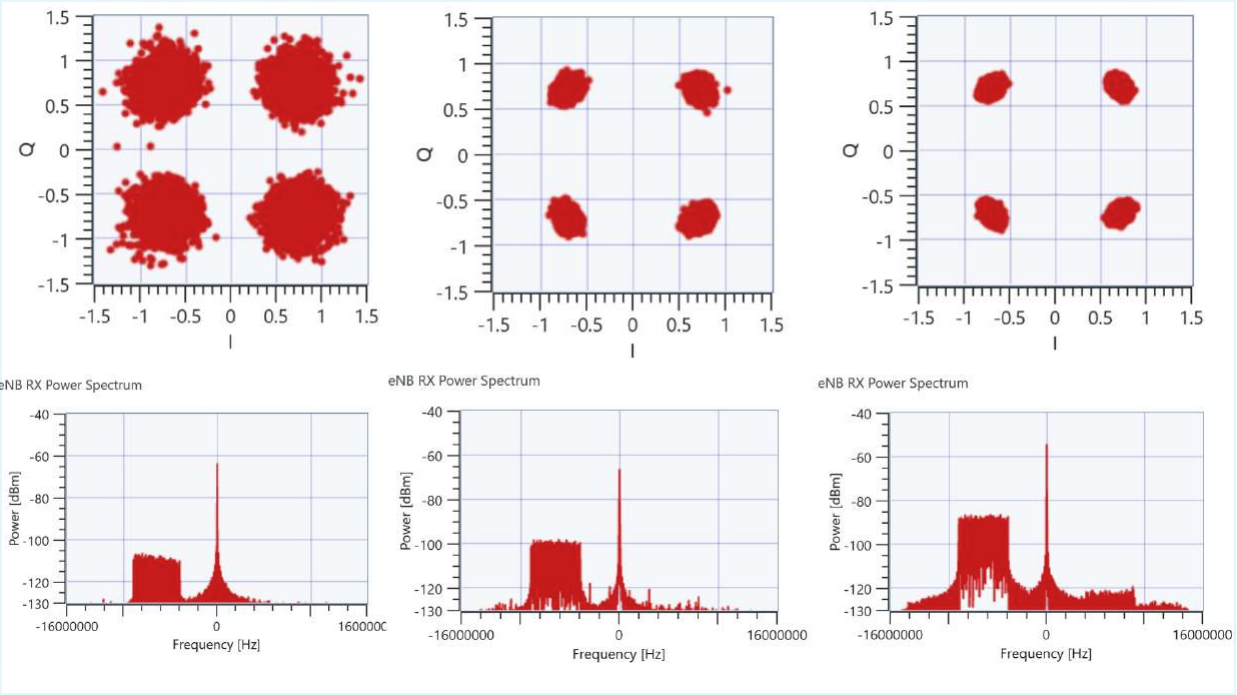}
    \put(17,0){\scriptsize (a)}
    \put(50,0){\scriptsize (b)}
    \put(84,0){\scriptsize (c)}
	\end{overpic}
	\vspace{2pt}
    \caption{Constellation diagrams (top) and received power spectra (bottom) at the eNB for a UE located at $45^{\circ}$ with respect to the RIS: (a) RIS OFF, (b) RIS all ON, and (c) RIS configured for $45^{\circ}$ beam steering.}
    \label{fig:constellation}
\end{figure}

\begin{figure}[!t]
	\centering
	\begin{overpic}[width=0.4\textwidth, height=0.43\textwidth, grid=false, tics=5, trim={0cm 0cm 0cm 0cm}, clip]{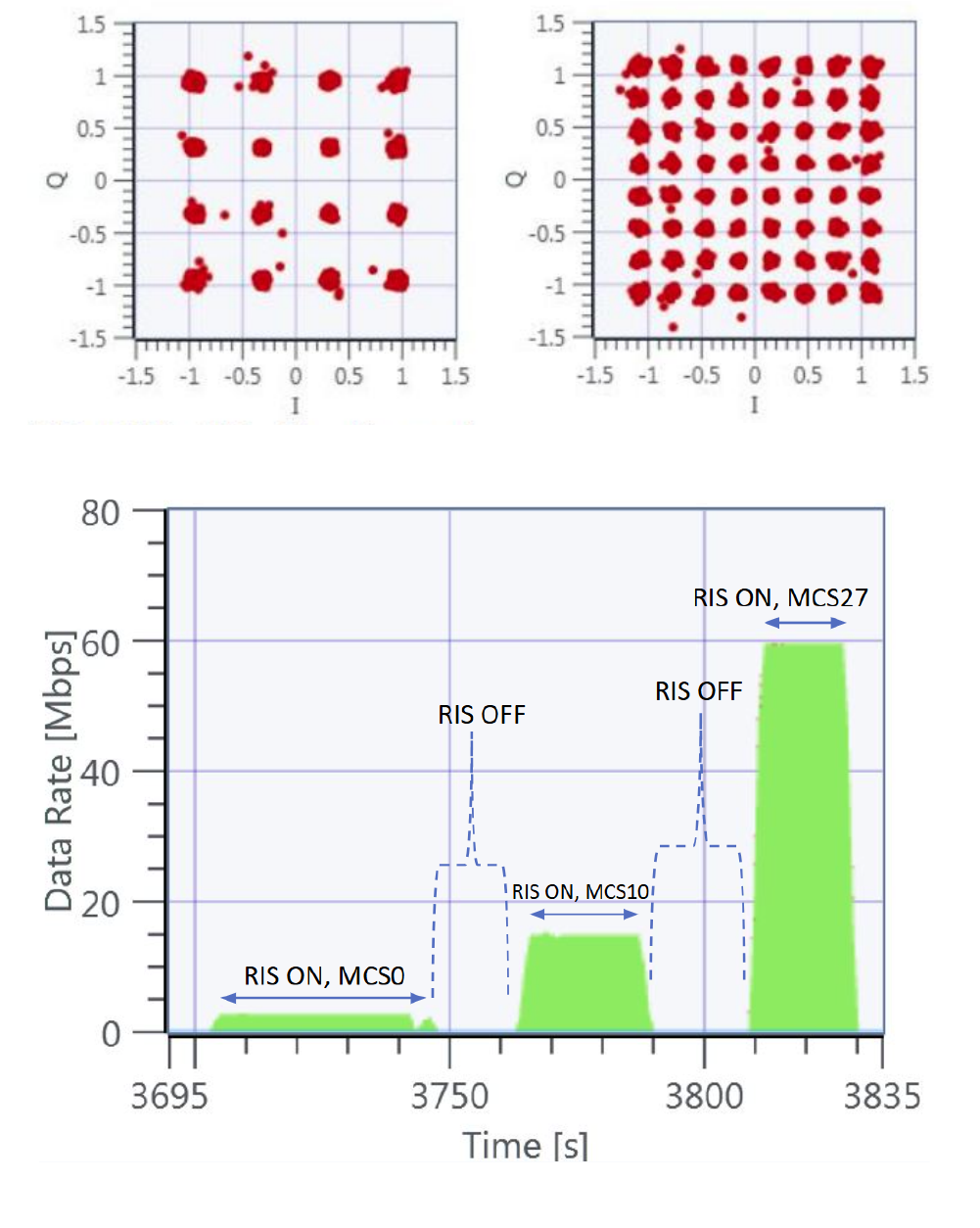}
    \put(27,62){\scriptsize (a)}
    \put(71,62){\scriptsize (b)}
    \put(45,0){\scriptsize (c)}
	\end{overpic}
	\vspace{2pt}
    \caption{Performance of the proposed RIS-assisted uplink across different modulation and coding Schemes (MCS). (a) 16-QAM constellation at MCS 10. (b) 64-QAM constellation at MCS 27. (c) Real-time throughput for different MCS when RIS is optimally configured and when it is OFF, respectively.}
    \label{fig:MCS}
\end{figure}

\subsection{Constellation and Spectral Characterization} 

Fig.~\ref{fig:constellation} illustrates the received signal characteristics at the eNB in terms of constellation diagrams (top row) and received power spectra (bottom row) for a UE located at an azimuth angle of $45^{\circ}$. The three columns correspond to different RIS operating states: (a) RIS with all PIN diodes OFF, (b) RIS with all PIN diodes ON, and (c) RIS configured for $45^{\circ}$ beam steering. When all PIN diodes are switched OFF [Fig.~\ref{fig:constellation}(a)], the received constellation exhibits significant symbol dispersion, indicating severe channel attenuation and low SINR. The corresponding power spectrum shows a weak uplink signal component close to the noise floor. Activating all RIS elements without directional phase optimization [Fig.~\ref{fig:constellation}(b)] results in moderate improvement, with more compact constellation clusters. However, spectral degradation due to insufficient Signal-to-Noise Ratio (SNR) remains visible due to the lack of focused beamforming. In contrast, when the RIS is configured to steer the reflected signal toward $45^{\circ}$ [Fig.~\ref{fig:constellation}(c)], the constellation points are tightly clustered around the ideal QPSK locations, indicating a substantially improved SINR. The corresponding received power spectrum shows an improved signal component with reduced spectral distortion due to the constructive combining from the RIS phase alignment.

To further evaluate the system's capacity under high-quality link conditions, the modulation and coding scheme (MCS) was manually varied within the LTE Application Framework. The resulting constellation diagrams and real-time throughputs are illustrated in Fig.~\ref{fig:MCS}. It can be observed that when the RIS is optimally configured for the $45^{\circ}$ steering direction, the obtained SINR can also successfully support higher-order modulation schemes. As shown in the figure, the data rate increases significantly from approximately 0.78 Mbps (MCS 0, QPSK) to 16 Mbps (MCS 10, 16-QAM), and then to over 60 Mbps at MCS 27 (64-QAM). So, even if our system’s rate adaptation algorithm automatically selects a lower-rate video stream, the RIS-aided link can provide higher throughput. In the 'RIS OFF' state, the constellation is completely collapsed, and throughput is near-zero. When RIS is optimally configured, the constellation points for all cases remain tightly clustered. This demonstrates that while the 1-bit quantization imposes a theoretical gain limit, the improved SINR due to the $16\times10$ array is sufficient to maintain the signal integrity required for high-throughput applications.

\section{Conclusion}
This work provides comprehensive details of a 1-bit coding RIS from unit cell design to its usefulness in UE localization and uplink communication. A uniquely designed single-polarized unit cell is used as the building block of the RIS. This unit cell can achieve a reflection phase difference of 180°±30°, along with a favorable magnitude of the reflection coefficient in the operating bandwidth of 22$\%$ on a low-cost FR4 substrate. The RIS, composed of a 16$\times$10 array of unit cells with varying coding patterns, provides beam steering in the xz plane for waves polarized along the y-axis. The system-level capabilities of the proposed RIS are demonstrated using a real-time LTE uplink testbed. This demonstration confirms that our cost-effective RIS hardware can reliably sustain high-throughput connectivity. Therefore, it can find many potential applications, such as UAV tracking and Industrial IoT in complex NLOS environments. Our future work will focus on mitigating the limitations of 1-bit RIS due to grating lobe, and scaling the array for outdoor deployments.

\balance
\bibliographystyle{IEEEtran}
\bibliography{References}

\end{document}